\documentclass[conference]{IEEEtran}

\usepackage{booktabs} % For formal tables
\usepackage{enumerate}
\usepackage{amsmath}
\usepackage[caption=false, font=footnotesize]{subfig}
\usepackage{amsfonts}
\usepackage{algorithm}
\usepackage{algorithmic}

\usepackage{listings}
\usepackage{xcolor}
\usepackage{caption}

\usepackage[english]{babel}
\usepackage{blindtext}
\usepackage{graphicx}
\usepackage{epstopdf}
\usepackage{etoolbox}
\usepackage{soul}
\makeatletter
\patchcmd{\maketitle}{\@copyrightspace}{}{}{}
\makeatother

\newcommand{\tabincell}[2]{\begin{tabular}{@{}#1@{}}#2\end{tabular}} 

\usepackage[normalem]{ulem}
\newcommand\redout{\bgroup\markoverwith{\textcolor{red}{\rule[.5ex]{2pt}{0.4pt}}}\ULon}

\def\mbf{\mathbf}
\def\mbs{\boldsymbol}
\def\mc{\mathcal}

\newcommand{\tens}[1]{\mc{#1}}

\usepackage{etoolbox}
\makeatletter
\patchcmd{\@makecaption}
  {\scshape}
  {}
  {}
  {}
\makeatother

\ifCLASSINFOpdf

\else
 
\fi

\begin{document}

\author{
    \IEEEauthorblockN{Bangtian Liu\IEEEauthorrefmark{1}, Chengyao Wen\IEEEauthorrefmark{1}, Anand D.  Sarwate\IEEEauthorrefmark{1}, Maryam Mehri Dehnavi\IEEEauthorrefmark{1}}
    \IEEEauthorblockA{\IEEEauthorrefmark{1}Rutgers, The State University of New Jersey
    \\\{bangtian.liu, chengyao.wen, anand.sarwate,  maryam.mehri\}@rutgers.edu}

}

%\title{From Matrix to Tensor: A Unified Optimization Approach for Sparse Tensor Operations on GPUs} 

\title{A Unified Optimization Approach for Sparse Tensor Operations on GPUs}

\maketitle

\begin{abstract}
Sparse tensors appear in many large-scale applications with multidimensional and sparse data. While multidimensional sparse data often need to be processed on manycore processors, attempts to develop highly-optimized GPU-based implementations of sparse tensor operations are rare. The irregular computation patterns and sparsity structures as well as the large memory footprints of sparse tensor operations make such implementations challenging. We leverage the fact that sparse tensor operations share similar computation patterns to propose a unified tensor representation called F-COO. Combined with GPU-specific optimizations, F-COO provides highly-optimized implementations of sparse tensor computations on GPUs. The performance of the proposed unified approach is demonstrated for tensor-based kernels such as the Sparse Matricized Tensor-Times-Khatri-Rao Product (SpMTTKRP) and the Sparse Tensor-Times-Matrix Multiply (SpTTM) and is used in tensor decomposition algorithms. Compared to state-of-the-art work we improve the performance of SpTTM and SpMTTKRP up to 3.7 and 30.6 times respectively on NVIDIA Titan-X GPUs. 
%\textcolor{blue}{
We implement a CANDECOMP/PARAFAC (CP) decomposition and achieve up to 14.9 times speedup using the unified method over state-of-the-art libraries on NVIDIA Titan-X  GPUs.%}

\end{abstract}

\IEEEpeerreviewmaketitle

\section{Introduction}
A tensor, a multi-dimensional or N-way array, represents multidimensional data naturally. 
Tensor-based computations or \textit{multilinear algebraic} methods such as tensor decomposition(s) appear widely in a variety of fields including machine learning~\cite{anandkumar2014tensor,huang2015online}, data mining~\cite{kang2012gigatensor,kolda2008scalable}, computer vision~\cite{vasilescu2002multilinear,vasilescu2011multilinear}, recommender systems~\cite{shi2012tfmap}, and quantum chemistry~\cite{khoromskaia2015tensor}. A number of industry-initiated frameworks for deep learning such as TensorFlow~\cite{abadi2016tensorflow} and Torch~\cite{collobert2011torch7} are also based on tensor representations.  Tensor operations are essential building blocks and tend to be the determinant operations for the performance of tensor algorithms and applications. In many applications, the tensors are sparse, that is,  most of their elements are zeros. Thus, developing parallel algorithms and libraries that accelerate sparse tensor computations on modern architecture is essential.

Previous work has optimized sparse tensor operations on different hardware platforms including shared memory systems ~\cite{smith2015splatt,smith2015csf,li2016optimizing}, distributed systems with MapReduce~\cite{kang2012gigatensor,jeon2015haten2} and on distributed memory with MPI~\cite{choi2014dfacto,kaya2016high,kaya2015scalable}. 
Due to their embarrassingly parallel execution model, GPUs are good candidates to accelerate sparse tensor computations; however, using GPUs is challenging because of the inherently irregular computation patterns in sparse tensor algebra. To our knowledge, Parallel Tensor Infrastructure (ParTI~\cite{jiajiaposter}) is the only work that accelerates tensor operations on GPUs. 
The optimizations in ParTI are not memory efficient, lead to load imbalance, and are sensitive to mode changes and increases in tensor computation ranks.
% \textcolor{blue}{
% However, ParTI consider sparse tensor operations as separate topics without unified view, what's more, 
% }

Previous work on tensor computations, optimize tensor operations independently and thus use different approaches to accelerate different sparse tensor operations. For example, the work in ~\cite{li2016optimizing,jiajiaposter,kaya2016high} optimizes the sparse
tensor-times-dense matrix (SpTTM) operation while others ~\cite{smith2015splatt,smith2015csf,choi2014dfacto,kaya2015scalable,jiajiaposter,kang2012gigatensor} mainly focus on the sparse Matricized Tensor Times Khatri-Rao Product (SpMTTKRP). The type of optimizations and the order to which they are applied are often shared between different sparse tensor operations. By investigating the underlying computation patterns and computation orders in sparse tensor operations, we propose an approach to generalize sparse tensor representations. Our unified  storage format and parallel algorithms can be used across many sparse tensor operations and can be extended to high-order tensor computations.  %Our unified representation enables us to introduce a number of GPU-specific optimization techniques for sparse tensor computations. 

Numerous challenges exist in optimizing sparse tensor operations on GPUs, such as \textit{i)} finding a good parallelization granularity, \textit{ii)} reducing storage costs and irregularities in memory accesses, and \textit{iii)} dealing with atomic updates. Prior work has used fiber- or slice-level computations as the granularity for parallelization. However, such approaches lead to noticeable load imbalance between threads on the GPU because of the sparse nature of the tensors. Also the optimizations do not deliver consistent performance for different modes and ranks.  The large intermediate data created during the sparse tensor computations is also very expensive to store on GPUs. Finally many of the sparse tensor operations require atomic updates that are expensive to perform on GPUs.
%This typically leads to load imbalance caused by the irregular sparsity patterns of the sparse tensor structures on GPU. Because of their irregular sparsity patterns in sparse tensor operations, {techniques such as memory coalescing for global memory accesses are also not directly applicable}. 
We propose a unified optimization method for sparse tensor operations to address these challenges on GPUs. Our major contributions are as follows:

\begin{enumerate}
\item \textbf{F-COO: A unified storage format for sparse tensors}. We propose a new storage format that is based on the tensor modes for sparse tensor computations. F-COO is memory efficient compared to other sparse tensor storage formats and can be used as a unified storage format across different sparse tensor operations.  
\item \textbf{Unified parallel algorithms for sparse tensor operations}. 
F-COO is used to propose parallel algorithms and optimizations for sparse tensor operations on GPUs. We demonstrate how optimizations of sparse tensor operations such as SpMTTKRP and SpTTM that have been treated differently in previous tensor literature are inherently the same.  Our unified parallel algorithms are used across different tensor operations, are  not sensitive to mode changes, and scale well with increases in the tensor computation rank. 
\item \textbf{GPU-specific optimizations}. By using the flag arrays in F-COO, we enable the application of efficient algorithms commonly used in sparse matrix literature such as the segmented scan method without unfolding the tensor. Other optimizations such as kernel fusion, warp shuffle, and data reuse are also enabled in our unified optimizations.% in the tensor computations. Through F-COO and an adaptation of the segmented scan algorithm~\cite{sengupta2008efficient,yan2013streamscan}, we propose unified optimizations for commonly used tensor operations to balance load across warps and reduce warp divergence, improve memory coalescing, and reduce memory access overheads on GPUs.  

\item \textbf{Significant speedups on real datasets for SpTTM, SpMTTKRP, and CP decomposition.} The proposed unified approach leads to $3.7\times$ speedup for SpTTM and $30.6\times$ speedup for SpMTTKRP for tested benchmarks over state-of-art work on GPU platforms. The CP decomposition is accelerated upto $14.9 \times$ times compared to state-of-the-art libraries.
Our unified method can be extended to support other sparse tensor operations and  other hardware platforms. 

\end{enumerate}

\section{Background}
%This section presents preliminary definitions and notations for sparse tensor operations used in the work.  ADS: not needed, takes up space
\noindent \textbf{Tensor notations:} A tensor is a multi-way array. The \textit{order} of a tensor refers to the number of dimensions, also called \textit{modes}. 
Vectors, first-order tensors, and matrices, second-order tensors, are presented by boldface lowercase and boldface capital letters receptively.
We generally use calligraphic letters for higher-order tensors (e.g., $\tens{X}$).
% A third-order tensor $\mathcal{X}\in\mathcal{R}^{I\times J \times K}$. 
The scalar element at position $(i,j,k)$ of a third-order tensor $\tens{X}$ is shown as $\tens{X}(i,j,k)$. We also use the colon notation from MATLAB (as does SPLATT~\cite{smith2015splatt}), in which a colon in the place of an index represents all members of that mode. For example, $\mbf{A}(m,:)$ is m-th row of the matrix $\mbf{A}$. 

% As shown in Figure \ref{fig:a}, a \textit{fiber} is a vector obtained from fixing every index but one. 
A \textit{fiber} is a one-dimensional segment of a tensor along one of the modes. A third-order tensor $\tens{X}$ has three kinds of fibers on three different modes represented by $\tens{X}(:,j,k)$, $\tens{X}(i,:,k)$ and $\tens{X}(i,j,:)$.
% As shown in Figure \ref{fig:b}, 
\textit{Slices} are two dimensional segments of a tensor, obtained by fixing all indices except for two. A third-order tensor $\tens{X}$ also has three kinds of slices, written as $\tens{X}(i,:,:)$, $\tens{X}(:,j,:)$ and $\tens{X}(:,:,k)$.

\begin{figure}
  \centering
  \includegraphics[width=0.25\textwidth,height=0.15\textwidth]{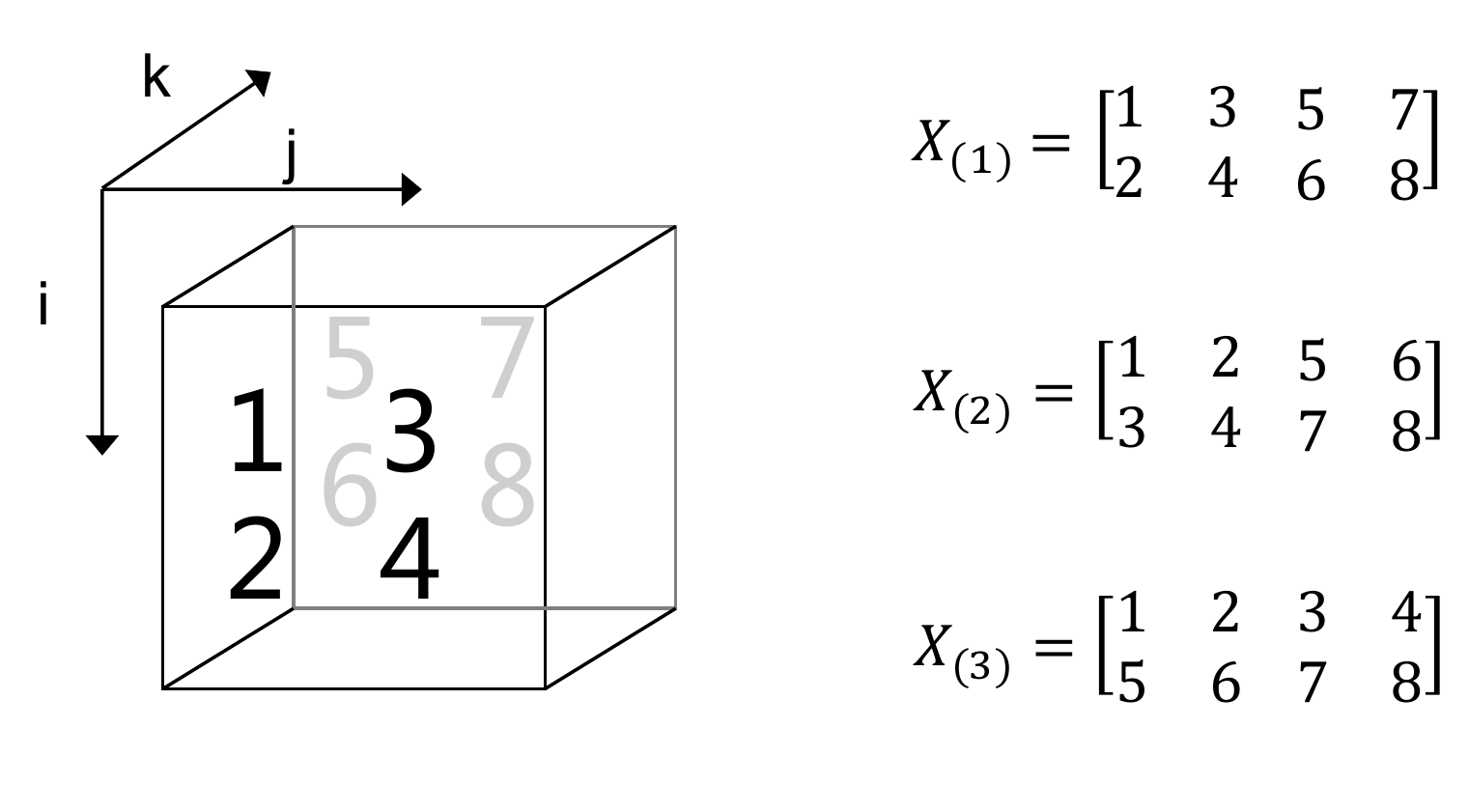}
  \caption{The matricization of a $(2\times2\times2)$ tensor.}
  \label{fig:unfolding}
\end{figure}

\textit{Matricization}, also called \textit{unfolding} or \textit{flattening},  transforms a tensor into a matrix.
The result of mode-$n$ matricization $\mbf{X}_{(n)}$ is a matrix which its columns are mode-$n$ fibers of the tensor $\tens{X}$ meaning that mode-n fibers in tensor $\tens{X}$ become the columns of the resulting matrix. Given a tensor $\tens{X}$ of size $I \times J \times K$, $\mbf{X}_{(1)}$ is of size $I\times JK$. Figure \ref{fig:unfolding} illustrates how to unfold a $(2\times2\times2)$ tensor along each of three modes. 
The \textit{Kronecker product} of matrices $A\in\mathbb{R}^{I\times J}$ and $B\in\mathbb{R}^{K \times L}$ is represented by $A \otimes B$, which generates a matrix of size $IK \times JL$. The Kronecker product is defined as
\begin{align}
A \otimes B = \begin{bmatrix}
			a_{11}\mbf{B} & a_{12}\mbf{B} & \dots  & a_{1J}\mbf{B} \\
            a_{21}\mbf{B} & a_{22}\mbf{B} & \dots  & a_{2J}\mbf{B} \\ 
			\vdots           &  \vdots          & \ddots & \vdots           \\     	
			a_{I1}\mbf{B} & a_{I2}\mbf{B} & \dots & a_{IJ}\mbf{B} 
            \end{bmatrix}
\end{align}

Given matrices $A$ and $B$, the \textit{Khatri-Rao product} of $A$ and $B$, also known as the \textit{column-wise Kronecker product}, is represented by $A \odot B$. %It requires that matrices $A$ and $B$ have same size in the dimension of column. 
If $A\in\mathbb{R}^{I \times K}$ and $B\in\mathbb{R}^{J \times K}$, the resulting matrix of size $IJ \times K$ is defined by 
\begin{align}
\mbf{A} \odot \mbf{B} = \begin{bmatrix}
			a_1 \otimes b_1 & a_2 \otimes b_2 & \ldots &	a_k \otimes b_k
			\end{bmatrix}
\end{align}

\textbf{Tensor-Times-Matrix: } Tensor-Times-Matrix (TTM) on mode $n$, also called the n-mode product, is a product of multiplying the tensor $\tens{X}\in\mathbb{R}^{I_1 \times I_2 \times \ldots\times I_n \times\ldots\times I_N}$ by the matrix $\mbf{U}\in\mathbb{R}^{I_n \times R}$ along the n-th dimension, represented by $\tens{Y}=\tens{X} \times_n U$. For third-order tensors, TTM on mode-3 is represented by:
\begin{equation}\label{eq:3TTM}
\tens{Y}(i,j,:)=\sum_{k=1}^{K}\tens{X}(i,j,k)\mbf{U}(k,:)
\end{equation}
When $\tens{X}$ is sparse and $\mbf{U}$ is dense, the operation is called SpTTM. In this case, the resulting tensor $\tens{Y}$ is only \textit{semi-sparse} since each fiber at index $(i,j)$ becomes dense and the length of the fiber will be equal to the number of columns of the dense matrix $\mbf{U}$. 
%\textcolor{black}{ 
SpTTM can be seen as a high dimensional generalization of the sparse matrix-vector multiply (SpMV) operation. SpTTM to tensor-based computation is what SpMV is to matrix-based computation. 
Similarly, SpTTM is usually the building block and bottleneck operation in tensor-based computations. 
%}

\textcolor{black}{
As demonstrated in \cite{kaya2016high,kolda2009tensor}, the key operation in the ALS-based Tucker decomposition algorithm, namely Higher
Order Orthogonal Iteration (HOOI), is called the tensor
times matrix-chain (TTMc) product. For an N-order tensor, TTMc on mode-n indicates tensor times matrix (TTM) products with $N-1$
different matrices along the corresponding modes other than mode-n. For example, a typical
mode-1 TTMc in Tucker decomposition for a 3-order tensor  is $\tens{Y}=\tens{X} \times_2 \mbf{U}_2 \times_3 \mbf{U}_3.$
Previous work on Tucker decomposition provides a high-performance parallel algorithm and implementation of TTMc~\cite{austin2015parallel,kaya2016high}. 
Equation \eqref{eq:ttmc} shows TTMc based on the coordinate storage format. More information about TTMc can be found in \cite{kaya2016high}.
}

 {
\begin{equation}\label{eq:ttmc}
\mbf{Y}_{(1)} = \sum_{\tens{X}(i,j,k)\in \tens{X}} \tens{X}(i,j,k) (\mbf{U}_2(j,:)\otimes\mbf{U}_3(k,:))
\end{equation}
}

%https://hal.inria.fr/hal-01354894/document 
\textbf{Matricized Tensor Times Khatri-Rao Product: }% Matricization of tensor and Khatri-Rao Product are two essential operations constituting Matriced Tensor Times Katri-Rao Product (MTTKRP).  For instance, e
MTTKRP is an important sparse tensor operation and is the main computation bottleneck in the CP decomposition algorithm. Parallel implementations of CP mainly focus on accelerating the execution of MTTKRP.
Equation \eqref{eq:mttkrp} represents MTTKRP operations along the first tensor mode. It unfolds the tensor along the first mode and then multiplies it with the Khatri-Rao product of the corresponding matrices $\mbf{B}$ and $\mbf{C}$:
\begin{equation}\label{eq:mttkrp}
M=\mbf{X}_{(1)}(\mbf{C} \odot \mbf{B})
\end{equation}
For large-scale sparse tensors, $\mbf{C} \odot \mbf{B}$ cannot be explicitly computed since the result matrix is dense and has a size of $JK \times R$ which can consume more memory than the sparse tensor $\tens{X}$ itself in most cases. Therefore, previous research on  SpMTTKRP focuses on how to map SpMTTKRP to other less-costly operations based on the sparsity pattern of the tensor to avoid computing the Khatri-Rao product explicitly~\cite{kang2012gigatensor,smith2015splatt}. The SpMTTKRP for the first mode can be written as follows:
\begin{align}
\mbf{M}(i,r) & = \sum_{z=1}^{JK}\mbf{X}_{(1)}(i,z)(\mbf{B}(z \% J,r)\mbf{C}(z/J,r))  \nonumber \\ 
\mbf{M}(i,:) & = \sum_{z=1}^{JK}\mbf{X}_{(1)}(i,z)(\mbf{B}(z \% J,:)*\mbf{C}(z/J,:)) \nonumber \\
				& = \sum_{k=1}^{K}\sum_{j=1}^{J}\tens{X}(i,j,k)(\mbf{B}(j,:)*\mbf{C}(k,:)), \label{eq:3mttkrp}
\end{align}
where $\%$ is the modulus operation.
While $\tens{X}$ is sparse, the result matrix $\tens{M}$ is a dense matrix where two product modes are replaced with the column dimensions of the dense matrices $\mbf{B}$ and $\mbf{C}$. The index mode $i$ is also dense due to the fact that a sparse tensor can not have empty slices in the $i$-dimension.

\textbf{Tensor Decomposition: }
The CANDECOMP/PARAFAC (CP) Decomposition factorizes a tensor into a sum of component rank-one tensors. The most popular algorithm for fitting the CP decomposition is based on the Alternating Least Squares (ALS) method. The ALS method iterates through the modes of the tensor, updating a factor matrix for each mode while holding the other factors constant. The algorithm for a 3-way tensor is given in Algorithm \ref{alg:3-way-CP-ALS}. 
\begin{algorithm}
	\caption{CP-ALS for a 3-way tensor}
    \label{alg:3-way-CP-ALS}
	\begin{algorithmic}[1]
		\REQUIRE {$\tens{X}$: A 3rd order tensor 
			R: The rank of approximation}
		\ENSURE CP decomposition $[\mbs{\lambda};\mbf{A},\mbf{B},\mbf{C}]$
		\REPEAT  
		\STATE $\mbf{A}\gets \mbf{X}_{(1)}(\mbf{C}\odot \mbf{B})                     (\mbf{B}^{\top}\mbf{B}\ast \mbf{C}^\top\mbf{C})^{\dagger}$ \label{line:1}\\ 
		\STATE Normalize columns of $\mbf{A}$ \\
		\STATE $\mbf{B}\gets \mbf{X}_{(2)}(\mbf{C}\odot \mbf{A})(\mbf{A}^{\top}\mbf{A}\ast \mbf{C}^\top\mbf{C})^{\dagger}$ \label{line:2}\\
		\STATE Normalize columns of $\mbf{B}$ \\
		\STATE $\mbf{C}\gets \mbf{X}_{(3)}(\mbf{B}\odot \mbf{A})(\mbf{A}^{\top}\mbf{A}\ast \mbf{B}^\top\mbf{B})^{\dagger}$ \label{line:3}\\
		\STATE Normalize columns of $\mbf{C}$ and store the norms as $\mbs{\lambda}$\\
		\UNTIL{no improvement or maximum iterations reached}  
	\end{algorithmic}
\end{algorithm}

\section{Related Work}

\textcolor{black}{
Sparse tensor operations such as SpTTM \cite{li2016optimizing,kaya2016high} and SpMTTKRP \cite{smith2015splatt, smith2015csf,choi2014dfacto} have been implemented as standalone routines to improve the performance of tensor algorithms and applications. Most of these work first propose or choose a tensor format and then propose parallel algorithms that operate on these formats efficiently. Therefore, our survey of previous work is based on \textit{(i)} storage formats for sparse tensors; \textit{(ii) }parallel algorithms on storage formats. 
}

\subsection{Storage Formats for Sparse Tensors}

While some parallel algorithms for sparse tensor operations are directly based on the coordinate format (COO) for SpMTTKRP~\cite{kaya2015scalable} and SpTTM~\cite{kaya2016high}, a number of novel storage formats have been proposed to reduce floating point computations and exploit more parallelism; the \textit{compressed data fiber} (CSF) format ~\cite{smith2015csf} is an example used in SpMTTKRP.  
These storage formats are proposed for distributed memory or shared memory systems. Dfacto~\cite{choi2014dfacto} and SPLATT~\cite{smith2015splatt} unfold a tensor along one mode to reduce floating point operations at the cost of increased memory usage. As pointed in~\cite{kaya2015scalable}, unfolding tensors requires column index values up to $\prod_{k \neq i}^{N}I_k$, which easily exceeds integer value limits when the input tensor is large in each mode.  
CSF is a tree-based data structure used in SPLATT that enables the extension of an efficient implementation of SpMTTKRP to higher dimensions. 
For GPU implementations of SpTTM on GPUs \cite{li2016optimizing}, Li \emph{et al.} proposes the semi-COO (sCOO) format which stores a semi-sparse tensor, which is the output of SpTTM. sCOO does not store indices for the dense modes in the semi-sparse tensor.

The objective of sparse tensor storage formats is to facilitate parallel implementations of tensor computations on shared and distributed memory systems. However, they often can not directly be used on GPU platforms since they will lead to significant overheads. For example, the \textit{compressed data fiber} (CSF) \cite{smith2015csf}, extends the compressed row storage format (CSR) to sparse tensors and is a fiber-centric and tree-based storage format. The recursive algorithms used in CSF-based optimization methods are not a good fit for GPU architectures. Also, the general storage format COO causes too many atomic operations when non-zeros processed by different threads share identical indices. The sCOO format stores the output of SpTTM which is a semi-sparse tensor~\cite{li2016optimizing}. Since sCOO is designed to store the output of a tensor operation, it can not be used to control the type of algorithms and optimizations used to implement the tensor operation.  

%In order to exploit parallelism from GPUs, we need a GPU-based storage format that is compatible with parallelizing sparse tensor operations. In this paper, we propose just such a unified storage format for sparse tensors.  We find commonalities between sparse tensor operations that lead to our new COO-based storage format. In addition to providing general support for sparse tensor operations on GPU, we show how to eliminate atomic operations on COO through a segment scan primitive supported by our proposed storage format. 

\subsection{Parallel Sparse Tensor Algorithms and Implementations}

Sparse tensor operations have been parallelized on different types of processing platforms such as shared-memory, distributed-memory, and GPUs.

%In this subsection, we introduce existing work on parallelizing and implementing sparse tensor operations on various platforms including shared-memory system, distributed memory system and GPU.
%
%
%
%

\textbf{Shared memory systems: }The Tensor Toolbox~\cite{bader2015matlab,kolda2008scalable} and N-way Toolbox~\cite{andersson2000n} are two widely used MATLAB toolboxes for tensor operations on share memory systems. The Cyclops Tensor Framework (CTF) \cite{solomonik2013cyclops} is a C++ library which provides automatic parallelization for sparse tensor operations. CTF transforms sparse tensors to matrices via unfolding and can only store the SpTTM output as a dense tensor. This restriction  significantly reduces its efficiency. SPLATT~\cite{smith2015splatt} is a library used for parallelizing the CP decomposition on shared-memory systems. It proposes a compressed and  fiber-centric data structure for sparse tensors called \textit{compressed data fiber} (CSF). Based on the CSF data structure, hypergraph models and multi-partite graphs are used to partition non-zeros into semi-sparse regions and improve data locality for SpMTTKRP.

\textbf{Distributed memory systems:} Many parallel algorithms have been proposed for large-scale tensor operations on distributed-memory systems.
Gigatensor~\cite{kang2012gigatensor} handles tera-scale tensors using the MapReduce framework. Gigatensor is the first work that minimizes the intermediate data sizes in SpMTTKRP and the number of floating operations for large-scale tensor operations. Dfacto~\cite{choi2014dfacto} also provides a distributed tensor decomposition implementation. However, the performance of Dfacto is limited by high memory footprints and data communication overhead since it needs to transform a tensor $\tens{X}$ into three matrices $\mbf{X}_{(1)}$, $\mbf{X}_{(2)}$, $\mbf{X}_{(3)}$ along three modes before operating on its data. Hypertensor~\cite{kaya2015scalable} is a sparse tensor library for SpMTTKRP on distributed-memory environments. Hypergraphs are used to partition the non-zero elements in a tensor and thus improve load balance and reduce data communication in sparse tensor operations on distributed memory environments~\cite{kaya2015scalable,kaya2016high}.

\textbf{GPU:} Li et al.~\cite{li2016optimizing} propose a parallel algorithm and implementation of SpTTM on GPUs, integrated in ParTI~\cite{jiajiaposter}, via parallelizing the algorithm on fibers. Since fibers in a sparse tensor may have different sizes their proposed implementation suffers from load imbalance and leads to warp divergence on GPU platforms for real sparse tensors. They also implement the SpMTTKRP algorithm on GPUs in ParTI~\cite{jiajiaposter} where data partitions are created based on the non-zeros of a tensor. The performance of their algorithm is limited by the overhead of atomic operations when updating divided slices.  
%Other than ParTI, there are few parallel algorithm and implementations on many-core GPU platforms.

\section{A unified optimization method for sparse tensor Operations on GPUs}

We propose a unified approach for the storage and optimization of tensor operations. We generalize some of the mode notations used in previous literature to categorize the computation patterns and structures in sparse tensor operations. The modes are then encoded into a novel sparse storage format that we call as F-COO (flagged-coordinate). F-COO can be used as a unified format across different tensor operations. We show how the mode encoding in F-COO allows our proposed parallel algorithms to operate on tensor non-zeros directly, eliminating the need to store intermediate data. This shows how a unified approach enables a one-shot approach to computing tensor operations such as SpMTTKRP. F-COO also enables the application of the segmented scan algorithm, a highly efficient algorithm used in sparse matrix computations, without the need to unfold the tensor into a matrix. In this section we describe our generalization of tensor modes, introduce F-COO, and show how this unified approach applies to parallel algorithms

\subsection{Unified Form of Sparse Tensor Modes}
The operations and computations in tensor methods can be characterized using a number of mode notations originally introduced by Li \textit{et al.} \cite{li2016optimizing} for SpTTM. In the following we extend these notations to other sparse tensor operations such as SpMTTKRP which enables us to propose a unified storage format and optimization method for sparse tensor operations: 
\begin{itemize}
\item \textit{Product modes}: are defined as the modes in which a tensor gets multiplied by a matrix. Mode-3 in Equation \eqref{eq:3TTM} for TTM and mode-(2,3) in Equation \eqref{eq:mttkrp} for MTTKRP are the product modes.
\item \textit{Index modes}: are all modes except for the product mode such as mode-(1,2) in Equation \eqref{eq:3TTM} for TTM and mode-1 in Equation \eqref{eq:mttkrp} for MTTKRP. 
\item \textit{Sparse mode}: is when at least one non-empty fiber in this mode is sparse. For example, if at least one of the fibers in $\tens{X}(i,j,:)$ is sparse mode-3 will be a sparse mode.
\item \textit{Dense mode}: is when all fibers in the mode are dense vectors. For example, if the fibers in $\tens{X}(i,j,:)$ are all dense then mode-3 is in a dense mode.
 \end{itemize}

\begin{table*}
\centering
\begin{tabular}{|c|c|c|c|c|c|c|c|}
\hline
Operations          &  Equation & Product mode & Index mode & Sparse mode of result & Dense mode of result \\
\hline
SpTTM on mode-3     & $\tens{Y}(i,j,:) \mathrel{+}=\tens{X}(i,j,k)\mbf{U}(k,:)$                   & mode-3 & mode-(1,2) & mode-(1,2) & mode-3\\
\hline
SpMTTKRP on mode-1  & $\mbf{M}(i,:) \mathrel{+}=\tens{X}(i,j,k)(\mbf{B}(j,:)*\mbf{C}(k,:))$  & mode-(2,3) & mode -1  & mode-1 & mode-(2,3)$\Rightarrow$mode-2\\
\hline 
SpTTMc on mode-1      &  $\mbf{Y}_{(1)}(i,:) \mathrel{+}= \tens{X}(i,j,k) (\mbf{U}_2(j,:)\otimes\mbf{U}_3(k,:))$     &    mode-(2,3) & mode-1 & mode-1 & mode-(2,3) \\     
\hline
\end{tabular}
\caption{Mode definitions for sparse tensor operations (mode-1: $i$, mode-2: $j$, mode-3:  $k$). The symbol $\Rightarrow$ indicates the mode change from input tensor to output. Based on mode classification, we can provide a unified view for sparse tensor operations. }\label{tb:unified}
\end{table*}

\subsection{The F-COO Storage Format}

This section discusses our proposed F-COO storage format which: \textit{i)} encodes changes in tensor modes and thus can be extended to support different sparse tensor operations; \textit{ii)} eliminates the need for tensor unfolding while enabling the application of efficient sparse matrix algorithms such as segmented scan; \textit{iii)}  requires less storage compared to formats used in previous tensor literature. %This section discusses the F-COO storage format and its application for different tensor operation.     % 
F-COO follows a similar storage approach to the COO format where all non-zeros of the tensor are stored with their corresponding indices and values. However, to enable unified computations for sparse tensor operations, the tensor modes discussed in the previous subsection are encoded into F-COO. As a result of this encoding, the F-COO captures changes in the computation pattern during the sparse tensor operations such as switching to a new fiber or slice or changing from a dense operation to a sparse mode. 

\begin{figure}
  \centering
\textbf{}    \includegraphics[width=0.5\textwidth,height=0.24\textwidth]{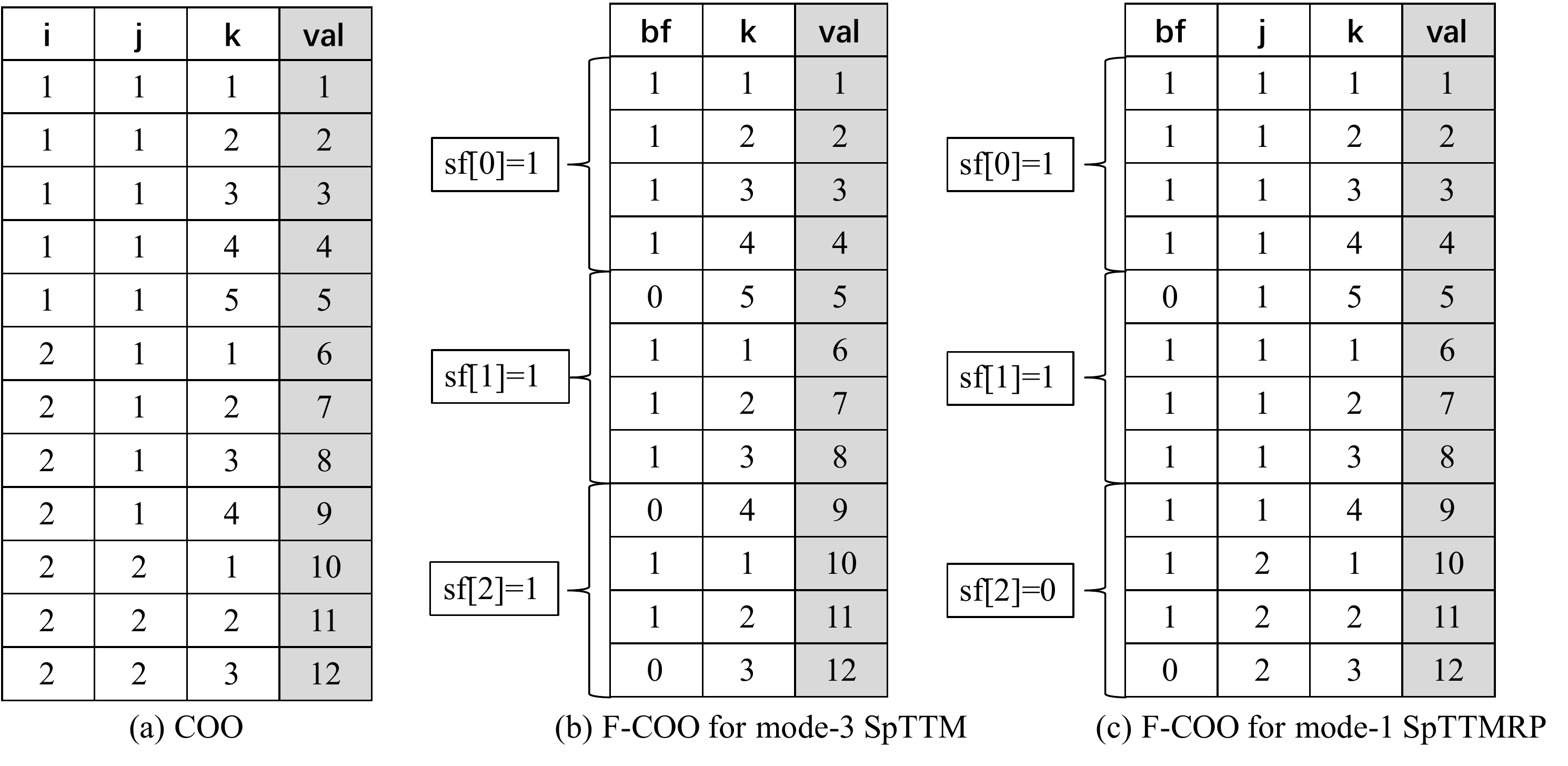}
  \caption{\textcolor{black}{
  F-COO for a 3-order tensor computing SpTTM on mode-3 and SpMTTKRP on mode-1. As shown in Table I for SpTTM,  the index modes are \textit{i} and \textit{j}.  bf (bit-flag) will change from 1 to 0  when a change in the \textit{i} or \textit{j} value occurs. For SpMMTKRP, bf change from 1 to 0 when the index mode \textit{i} changes. 
If each partition holds 4 tensor non-zeros, sf (start-flag) indicates whether partitions processed by the current thread start new indices on index mode over previous thread. sf for thread 0 is always 1 since it always starts new indices.
  }
  }

  \label{fig:storageformat}
\end{figure}

Table \ref{tb:unified} shows our classification of tensor modes for different operations and Figure \ref{fig:storageformat} demonstrates how this classification is used to store the tensor for the SpTTM and SpMTTKRP operations. The $val$ vector stores the non-zero values of the tensor. Except for the flag arrays, all other vectors such as $i, j, k$ are used to store the indices corresponding to the product mode. The indices in F-COO that correspond to the product mode are used to guide the Kronecker or Hadamard product operations. F-COO also uses two flag arrays, i.e. the bit-flag (bf) and the start-flag (sf). The bf array is used to represent any changes in the index modes which consequently shows the computation has switched to another fiber (in SpTTM) or to another slice (in SpMTTKRP). F-COO also comes with a start-flag (sf) that is used to indicate whether a new fiber or slice starts inside the current partition. Section IV-D demonstrates how the flag arrays are used to implement segmented scan to remove atomic updates and increase parallelism in tensor computations. 
Table I and the F-COO storage format can be extended to support other tensor operations and higher-order tensors.

As demonstrated in Figure \ref{fig:storageformat}, F-COO is used as a unified storage format for different tensor operations reducing tensor storage costs and enabling the application of unified parallel algorithms across tensor operations. Existing methods optimize tensor operations in isolation, requiring a different storage format and optimization strategy for each tensor operation~\cite{jiajiaposter}. 
For example, ParTI (the only work that accelerates sparse tensor operations on GPUs) 
%which to our knowledge is the only work that accelerates sparse tensor operations on GPUs 
parallelizes SpTTM on the tensor fibers where the input is stored in a compressed fiber-order. For the SpMTTKRP operation, ParTI uses the COO storage format to enable operating on the non-zeros of the tensor. Our unified algorithm and storage format captures the similarity between these two operations; the F-COO storage format allows efficient operations on tensor non-zeros for both operations and is significantly faster than ParTI.

%In this paragraph, we present how to formalize F-COO for sparse tensor operation in detail. 
%As shown in Figure \ref{fig:storageformat}, for SpTTM on mode-3, F-COO only keeps $k$ array on the product mode (mode-3) and the data value $val$ array  on the GPU, using bit flag array to replace the index arrays on the index mode (mode-(1,2)).  Beside flag array, 
%\textit{startflag} array
% is used to support segmented scan primitives for CUDA threads and guarantee the correctness of the parallel algorithm to sum divided fiber or slice spanning on multiple threads. The type of flag arrays is determined by the number of non-zeros processed per thread. 

Like COO, F-COO stores non-zero tensor elements. However, it does not suffer from load imbalance and can maintain maximum parallelism when operating on sparse tensors on different modes. Also, similar to COO, F-COO is insensitive to the irregularities of the underlying sparse tensor structures; this is why COO is useful in sparse matrix computations~\cite{bell2009implementing}. One of the major drawbacks of using COOs in tensor computations is that COO has a high memory footprint because all the product and index mode indices have to be explicitly stored and accessed. Compared to COO, F-COO is more memory-efficient because it only keeps the indices on the product mode; the index modes are not stored and only a change in their values are stored in a considerably smaller bit-flag array. The number of non-zeros processed per thread depends on the data type selected for the bf array shown in Figure \ref{fig:storageformat}. For example, if we use uint8\_t or unsigned char to bf, the number of non-zeros processed per thread will be 8.
% and thus a bit-flag array of type uint8\_t or unsigned char  is used to compress 8 bits for non-zeros accessed by threads. 
For the sf array, we use unsigned int to compress 32 bits for values accessed by threads in one warp concurrently to save bandwidth. Table II compares the storage costs of COO and F-COO for the SpTTM and SpMTTKRP operations; these can be extended to other sparse tensor operations.
  We do not compare to CSF \cite{smith2015csf} because it is for CPU use only.

\begin{table*}
\begin{tabular}{|c|c|c|}
\hline
storage format & SpTTM on mode-3 &  SpMTTKRP on mode-1 \\ 
\hline
COO          & $16 \times nnz$  &   $16 \times nnz$  \\
\hline
F-COO        &  $(8+1/8+1/(8*threadlen))\times nnz$  &   $(12+1/8+1/(8*threadlen))\times nnz $     \\
\hline
\end{tabular}
\centering
\caption{
Storage cost of a 3-order tensor for COO vs. F-COO. The storage cost of COO for a 3-order sparse tensor is $16\times nnz $ bytes when integer and single-precision floating-point are used to store indices and non-zeros respectively. For SpTTM, indices in the index mode are replaced by bits, therefore, F-COO for SpTTM only takes $(8+1/8+1/(8*threadlen))\times nnz$ bytes   (\textit{threadlen} indicates the number of non-zeros processed per thread), where $8 \times nnz$ bytes are the storage cost for indices in the product mode (one integer per non-zero) and data values (one float per non-zero), $(1/8)nnz$ bytes is the memory cost from the bit-flag array and $1/(8*threadlen)nnz$ is the storage cost for the start-flag array, which is a flag array for each thread. The storage cost of SpMTTKRP is obtained analogously. 
% Variable \textit{threadlen} indicates how many non-zeros processed by each thread. 
% For example, if sparse tensor has 16 non-zeros and each thread processed 8-zeros, we will issue 2 threads to parallelize the computation.
% The upper bound and upper bound depends the type we can used to encode bit flag in F-COO. 
% The type provided by programming language is char (8), short (16), int (32), long(64). 
% Therefore, the option of \textit{threadlen} is 8, 16, 32 and 64. 
}\label{tb:cost}
\end{table*}

\subsection{One-shot sparse tensor computations}
\begin{figure*} 
\centering 
\subfloat[Previous method for SpMTTKRP] { \label{ofiber} 
\includegraphics[width=0.26\textwidth, height=0.3\textwidth]{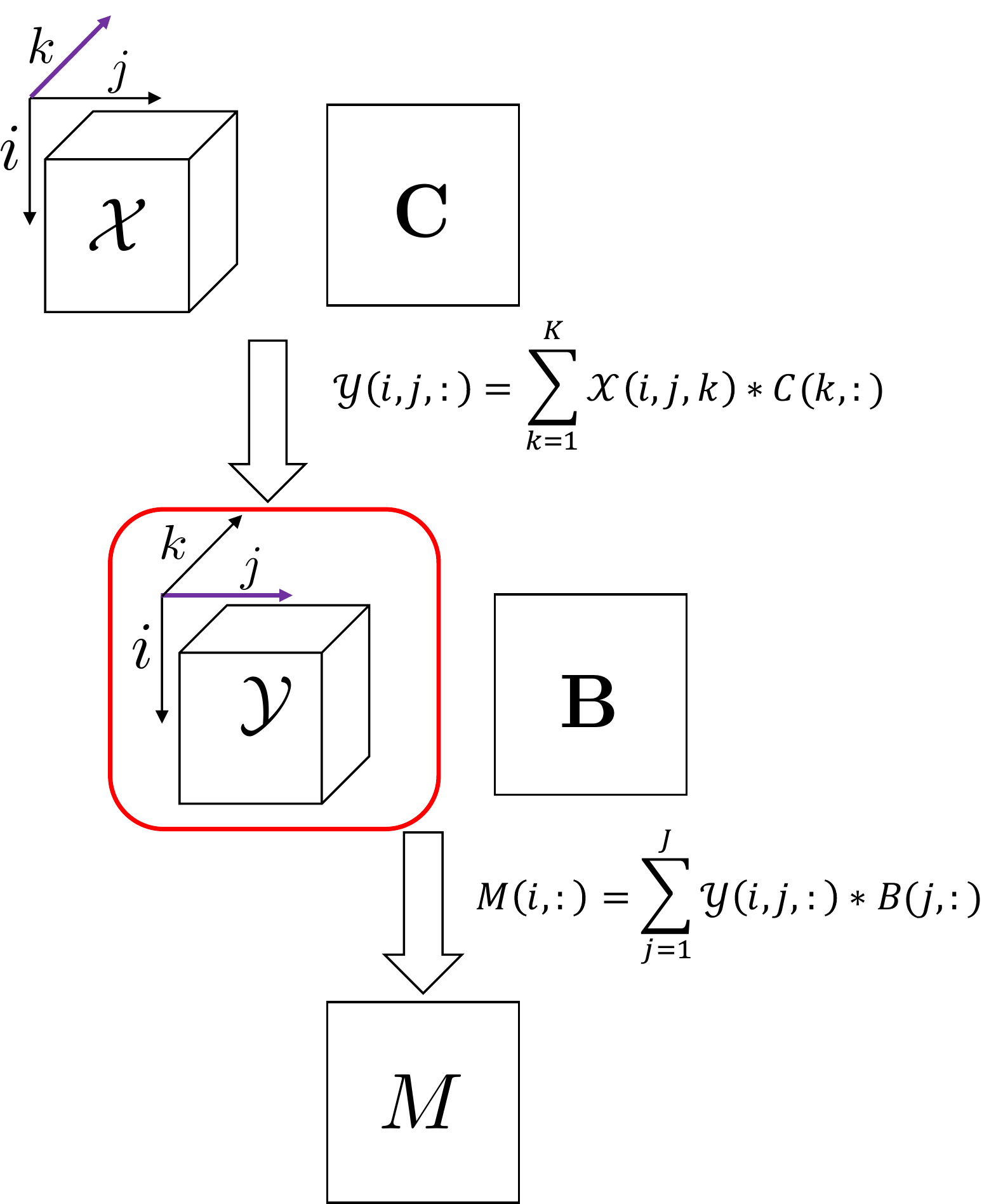} 
} 
\centering
\subfloat[One-shot method for SpMTTKRP ] { \label{oneshot} 
\includegraphics[width=0.26\textwidth, height=0.23\textwidth]{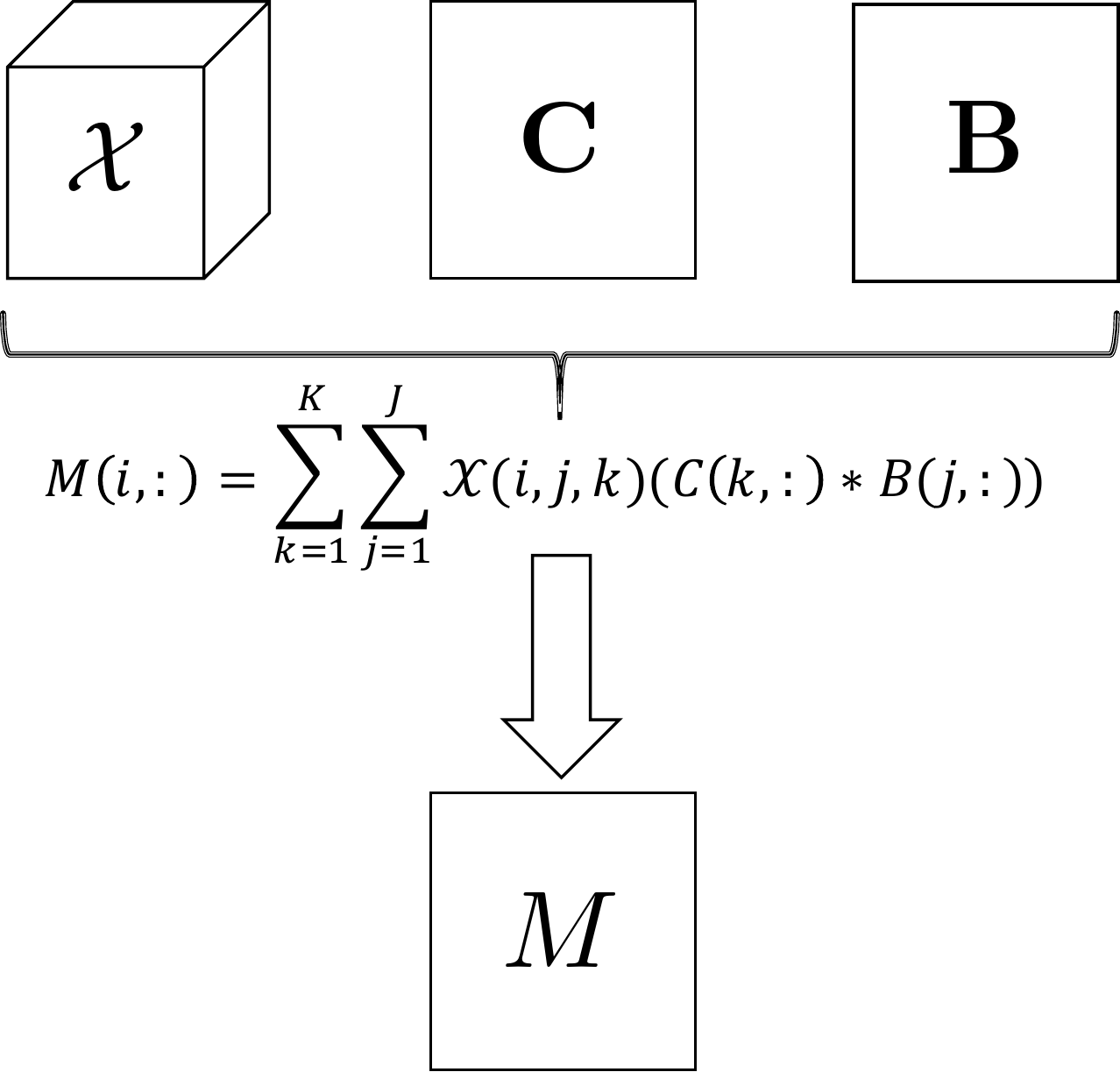} 
} 
\caption{Figure (a) shows that previous fiber-centric SpMTTKRP implementations  first multiply along mode-k with matrix C, then multiply along mode-j with matrix B. The drawback is that an intermediate tensor $\tens{Y}$ is generated and tensor operations will have to switch between different modes. 
Figure (b) illustrates our proposed one-shot method for SpMTTKRP. Our method directly performs computations on the non-zeros of the sparse tensor in one-shot.   
} \label{oneshotcom}
\label{fcomparison}
\end{figure*}

In the proposed unified approach, the F-COO storage format is used to compute tensor operations in \textit{one-shot}. The one-shot strategy eliminates the need to create large intermediate data and avoids transformations between different F-COO representations.
%Sparse tensor operations represented by F-COO should be performed in one-shot on GPU for two reasons: \textit{(i)} significant overhead in converting between different F-COO representations for corresponding tensor operations; \textit{(ii)} large intermediate data imposes storage pressure on limited size of GPU memory. In the following paragraphs, we elaborate how we manage to achieve one shot sparse tensor operation on GPU. 
As shown in Figure \ref{ofiber}, when a sparse tensor operation such as SpMTTKRP is transformed into a series of sparse computations it will generate intermediate tensors which lead to extra storage costs. Figure \ref{ofiber} shows that operating on the sparse tensor $\tens{X}$ of size $I \times J \times K $ will generate an intermediate tensor $\tens{Y}$ of size $I \times J \times R $, which has larger storage costs compared to $\tens{X}$ because mode-3 is dense. Also, operations on intermediate tensors require mode change in F-COO which is an expensive operation.

As shown in Figure \ref{oneshot}, the overhead in both storage cost and conversion between different F-COO representations is eliminated when performing sparse tensor operations in one-shot. 
The one-shot algorithm uses the product mode indices in F-COO to obtain rows from the dense factor matrices, C or B, and to perform a Hammard or Kronecker product. This intermediate result is then scaled using the corresponding non-zero value in the sparse tensor and is accumulated to the correct location using the indices from the index mode. The results are accumulated using segmented scan to reduce atomic operations in the parallel implementation.

\subsection{Parallel algorithms for sparse tensor operations on GPUs}

\textcolor{black}{
We propose unified parallel algorithms for sparse tensor operations on GPUs based on the F-COO storage format. This section discusses how our unified algorithm uses the segmented scan primitive to improve parallelization and how it reduces atomic updates in sparse tensor computations. We will also discuss our parallelization strategy which operates on the non-zeros of the sparse tensor to maintain load balance and provides consistently good performance for different ranks in sparse tensor computations. GPU-specific optimizations techniques are also discussed. Finally, we will demonstrate how unified can be used to implement complete tensor algorithms such as CP decomposition.
}

\textbf{Enabling segmented scan:
} Using the F-COO storage format, the non-zeros in the tensor are accessed to apply the computations in parallel and reduce the results using the product indices. 
The F-COO format has two flags, the bit-flag and the start-flag, both of which are used to implement the segmented scan algorithm to parallelize and reduce product results in the sparse tensor computation. The bit-flag is toggled when a new fiber or slice starts and the start-flag is used to indicate the start of a fiber or slice inside a partition. Partitions are allocated to different threads and their sizes are tuned for best performance. Details of the segmented scan algorithm can be found in \cite{sengupta2008efficient,yan2013streamscan} and are not repeated here. %With flag arrays in F-COO, We then perform a segment scan primitive to sum values across threads and obtain the final correct result without using atomic operations. 
% Algorithm \ref{alg2} just computes partial result of sparse tensor operations on GPU. 
%Take SpMTTKRP as an example, we firstly computes the appropriate $\tens{X}(i,j,k)(\mbf{B}(j,:)*\mbf{C}(k,:))$ product, Then we will add corresponding values of same row (i) across threads and blocks with the segmented scan algorithm, which is usually used for SpMV ~\cite{yan2014yaspmv,bell2009implementing}. 

\begin{figure*}[!htp]
  \centering
    \includegraphics[width=0.6\textwidth,height=0.32\textwidth]{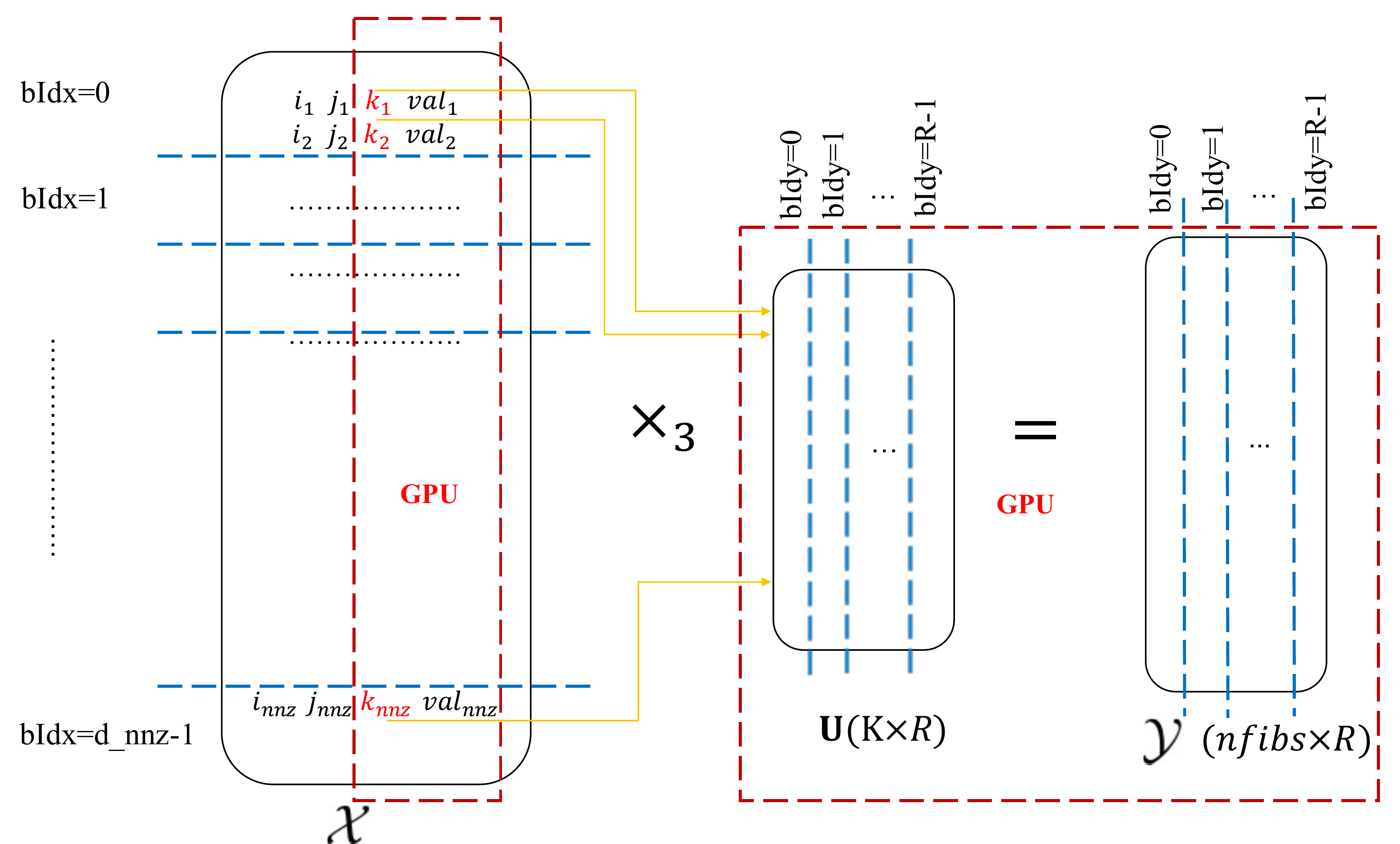}
  \caption{The parallelization and partitioning strategy used in unified for SpTTM on mode-3 ($(\tens{Y}=\tens{X}\times_3U)$). R is the number of columns in the dense factor matrix. 
  %Our parallel implementation launches two-dimensional thread grids with one-dimensional thread blocks.
  A thread block is shown by a two-dimensional index (bIdx, bIdy) in the figure. $d\_{nnz}$ represents the dimension of the thread grid along the \textit{x} dimension: non-zeros of sparse tensor, R represents the dimension of thread grid along the \textit{y} dimension: column dimension of the dense matrix.  
  }
  \label{gpupartition}
\end{figure*}

\textbf{Parallelization strategy:
} The unified approach partitions the data and parallelizes computations based the non-zeros of the sparse tensor and the columns of the dense factor matrices as shown in Figure \ref{gpupartition}. As a result, our approach delivers consistently good performance  for larger factor matrices  where the rank of the tensor operation increases.   The number of columns in dense matrices represents the rank of tensor decomposition. 
Previous work on sparse tensor optimizations on GPUs \cite{li2016optimizing} uses two-dimensional thread blocks in their implementation where the shape of thread blocks varies with the rank of the tensor operation. For example, when the number of threads is 512 in a two-dimensional thread block and rank is 32, the shape of the two-dimensional thread block will be $(16,32)$. Since the threads inside a warp are in charge of computing the product of two dense columns in the implementations proposed in  \cite{li2016optimizing}, the shape of the thread block can lead to thread divergence inside a warp and cause strided memory accesses. As a result, the performance of the code from  \cite{li2016optimizing} can vary for different ranks of the tensor operation.

To resolve this issue, we launch two-dimensional thread grids with one-dimensional thread blocks.  One-dimensional thread blocks operate on their allocated partitions of the sparse tensor and columns of the dense matrices indicated by thread block index (bIdx, bIdy) as shown in Figure \ref{gpupartition}. Since the thread block dimensions in our approach do not vary with rank, the rank of sparse tensor operations will not affect parallelism and the memory access patterns in the proposed unified approach.

% \textbf{Parallelization strategy:
% } Unified method obtains parallelism from both non-zeros of sparse tensor and columns of dense matrix as shown in Figure \ref{gpupartition}. The number of columns of dense matrix in sparse tensor operations represent the rank of tensor decomposition, whose essential building blocks usually are constituted by these sparse tensor operations. 
% \edit{To avoid the parallelism and memory access pattern of each thread block affected by the Rank}{In case that the parallelism and memory access pattern of each thread block vary for different Rank---number of columns of dense matrices}, Our choice is to organize two-dimensional grid of one-dimensional thread blocks. As shown in Figure \ref{gpupartition}, one dimension of two-dimensional thread grid is determined by non-zeros, another dimension is determined by the number of columns of dense matrix.  
% For example, if we set a two-dimensional thread blocks as ParTI \cite{li2016optimizing} for SpTTM, when the number of threads is 512 in two-dimensional thread block and Rank is 32, the shape of two-dimensional thread block is $(16,32)$.
% In this case, the number of thread in x dimension is 16 ($<32$). 
% To avoid the shape of thread block is affected by Rank, we organize thread blocks in one-dimension and organize the grid into two dimension based on \edit{the Rank}{non-zeros of sparse tensor and column dimension of dense matrices}.   

\textbf{GPU-specific optimizations:}
A number of GPU-specific optimization techniques are used to further improve the performance of our unified algorithm on GPUs. Since sparse tensor operations on GPUs are memory-bound all of the techniques are memory-oriented optimizations to efficiently use the GPU memory hierarchy. Our optimizations include using the read-only data cache, fusing kernels, and applying warp shuffle. 
%
%Since each thread will access input tensor and matrix with a stride simultaneously, we will perform off-line transpose for dense matrix and input tensor to avoid stride memory access.
Since in a single SpTTM and SpMMTKRP operation the dense factor matrices are read-only, they are cached in the Read-Only Data-Cache to further reduce global memory loads. Adjacent synchronization~\cite{yan2013streamscan} is used to perform inter-block communication and to fuse the kernels in the sparse tensor implementation. Kernels such as the product kernel, segmented scan, and the accumulation kernels are fused to increase data reuse and keep intermediate data in shared memory.
For the segment scan implementation, warp shuffle is used to increase data sharing inside a warp. Warp shuffle enables register to register data exchange and thus reduces the shared memory footprint and avoids overusing shared memory.

\textbf{Complete tensor-based algorithms:
} Sparse tensor operations such as SpMTTKRP and SpTTM are used inside \textit{complete tensor-based algorithms} such as the Tucker and CP decomposition algorithms. To our knowledge there are currently no implementations of CP or Tucker for sparse tensors on GPUs. We implement the CP decomposition algorithm to show our unified approach is insensitive to the mode being operated on. As a result, the MTTKRP operations in lines 2, 4, and 6 in Algorithm 1 will have very similar and well-balanced execution times.  To eliminate the need for format conversations or CPU-GPU data transfers inside a CP iteration, F-COO is preprocessed for different modes on the host and will only be transferred once in the beginning to the GPU. For very large tensors, multiple-GPUs can be used. A similar approach can be used to implement Tucker using unified.

% \textcolor{black}{
% For example, as shown in Algorithm \ref{alg:3-way-CP-ALS}, 
% SpMTTKRP operations on different modes (line \ref{line:1}, line \ref{line:2} and line \ref{line:3}.) constitute a complete CP decomposition algorithm. In this Algorithm, all the format of intermediate result is known before running the algorithm. Therefore, F-COO for SpMTTKRP operation on different modes is prepared in advance on host and then transfer the F-COO data to device.   
% }

% \textcolor{black}{Bangtian: 
% Another concern may be whether GPU memory is not large enough to accommodate all F-COO on different modes. 
% In the worst case, the solution is  we can overlap the matrix computation at line \ref{line:1} and memory transfer resulted by SpMTTKRP \ref{line:2}, after finishing SpMTTKRP operations.
% }

\section{Experiments}
\label{sec:experiments}

We evaluate the performance of the unified approach by comparing to two state-of-the-art tensor libraries, namely ParTI \cite{jiajiaposter} and SPLATT \cite{smith2015splatt}. %\subsection{Reference Algorithms and Implementations}
ParTI accelerates sparse tensor operations on multicore CPU and GPU architectures.  SPLATT provides high-performance implementations of SpMTTKRP on shared-memory systems. SPLATT doesn't support sparse tensor operations on GPUs. %These two libraries are currently known as the state-of-the-art work on sparse tensor operations. 
%ParTI~\cite{jiajiaposter} and SPLATT libraries are used as the reference scenario to evaluate the performance of unified method proposed in this paper. 
All the experiments on the CPU platform for ParTI-omp and SPLATT are executed with 12 threads. % including ParTI-omp and SPLATT on shared-memory system. 
For fair comparison, we follow the execution instructions provided by the authors of SPLATT and ParTI libraries and thank them for their assistance in this process.    

 \begin{table}[!htp]

\begin{tabular}{ c c c }
\toprule
  Parameters & \tabincell{c}{Intel \\ Core i7-5820K} & \tabincell{c}{NVIDIA \\ GeForce GTX Titan X}   \\
  Microarchitecture & Haswell & Maxwell \\
  Frequency& 3.3GHz & 1.0 GHz \\
  Physical cores & 6 & 3072 \\
  Peak SP Performance &56.72 Gflops&6144Gflops \\
 \midrule
 Last-level cache & 15MB& 3MB\\
 Memory size & 64GB &12GB \\
 Memory bandwith &68 GB/S &336GB/S \\
 \midrule
 compiler & gcc 5.4.0 & nvcc 8.0\\
 \bottomrule 
\end{tabular}
\caption{Experimental platform configuration.}\label{platform}
\end{table}

\begin{table}[!htp]

\begin{tabular}{ccccc}
\toprule
Dataset & order & Mode sizes                     & nnz  & density \\
\midrule
brainq  &  3    & $60 \times 70K \times 9 $       & 11M  &  $2.9e-01$ \\
nell2   &   3   & $12K \times 9K \times 29K$     & 77M  &  $2.5e-05$  \\ 
dellicious & 3  & $0.5M\times 17.3M \times2.5M $ & 140M & $6.1e-12$ \\
nell1      & 3  & $2.9M\times 2.1M \times 25.5M$ & 144M & $9.3e-13$ \\
\bottomrule
\end{tabular}
\caption{Description of sparse tensor datasets.}\label{dataset}
\end{table}
 
%\subsection{Experimental Setup}

Our experiments are performed on an Intel Core i7-5820K CPU and the NVIDIA GeForce GTX Titan X platforms. Hardware configurations are shown in Table \ref{platform}. 
We use a number of datasets from real applications provided by FROSTT~\cite{frosttdataset}. Nell1 and nell2 come from Never Ending Language Learning (NELL) project~\cite{carlson2010toward} and represent \textit{noun-verb-noun} triplets. The brainq dataset is generated from functional Magnetic Imaging (fMRI) measurements of brain activity~\cite{mitchell2008predicting}, which represents \textit{noun-voxel-human} subjects. Delicious is a \textit{user-item-tag} tensor crawled from tagging systems~\cite{gorlitz2008pints}. Table \ref{dataset} provides more detail on the datasets.

\begin{figure*} 
\centering 
\subfloat[brainq] { \label{brainq3d} 
\includegraphics[width=0.4\textwidth, height=0.24\textwidth]{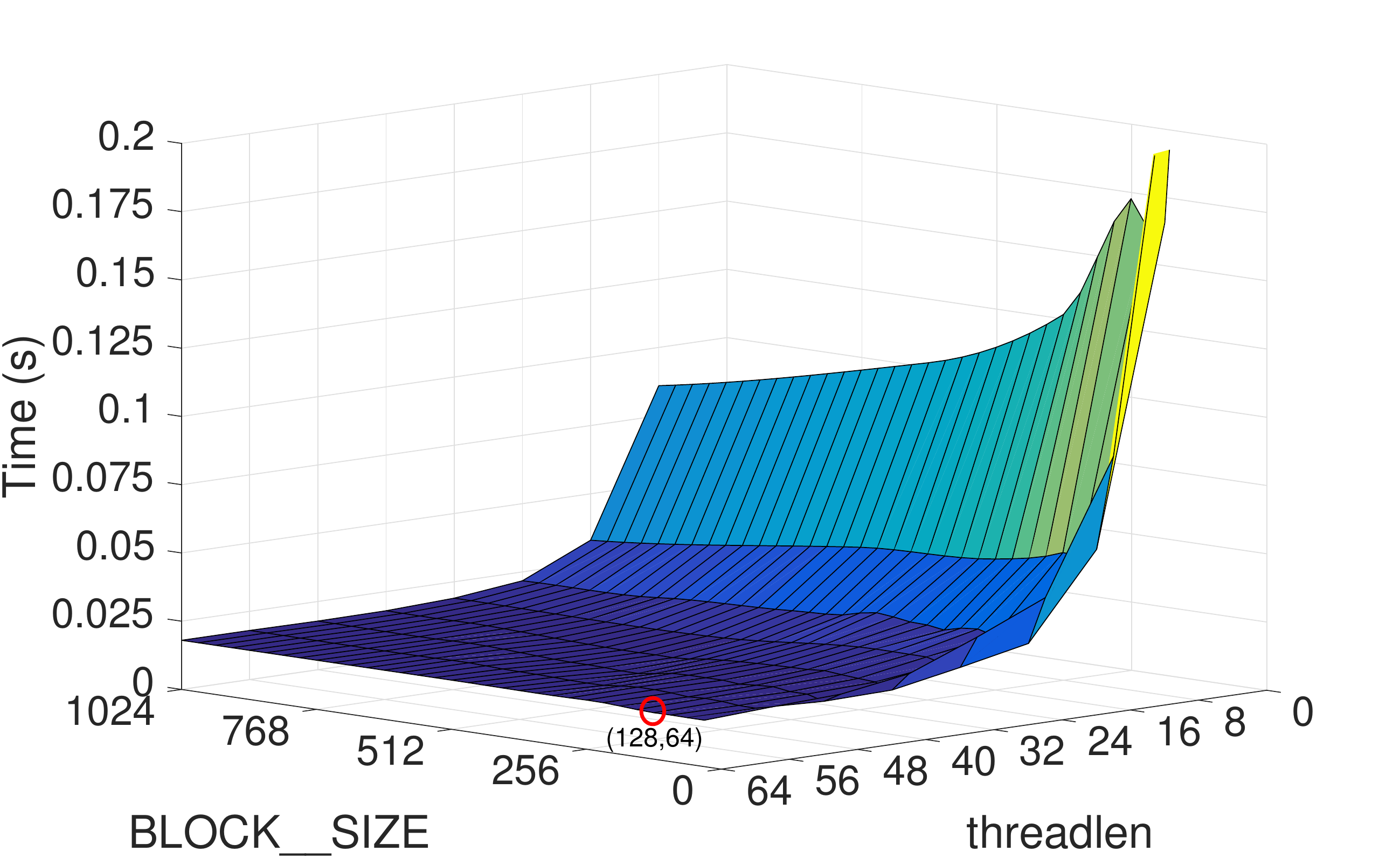} 
} 
\centering
\subfloat[nell1] { \label{nell13d} 
\includegraphics[width=0.4\textwidth, height=0.24\textwidth]{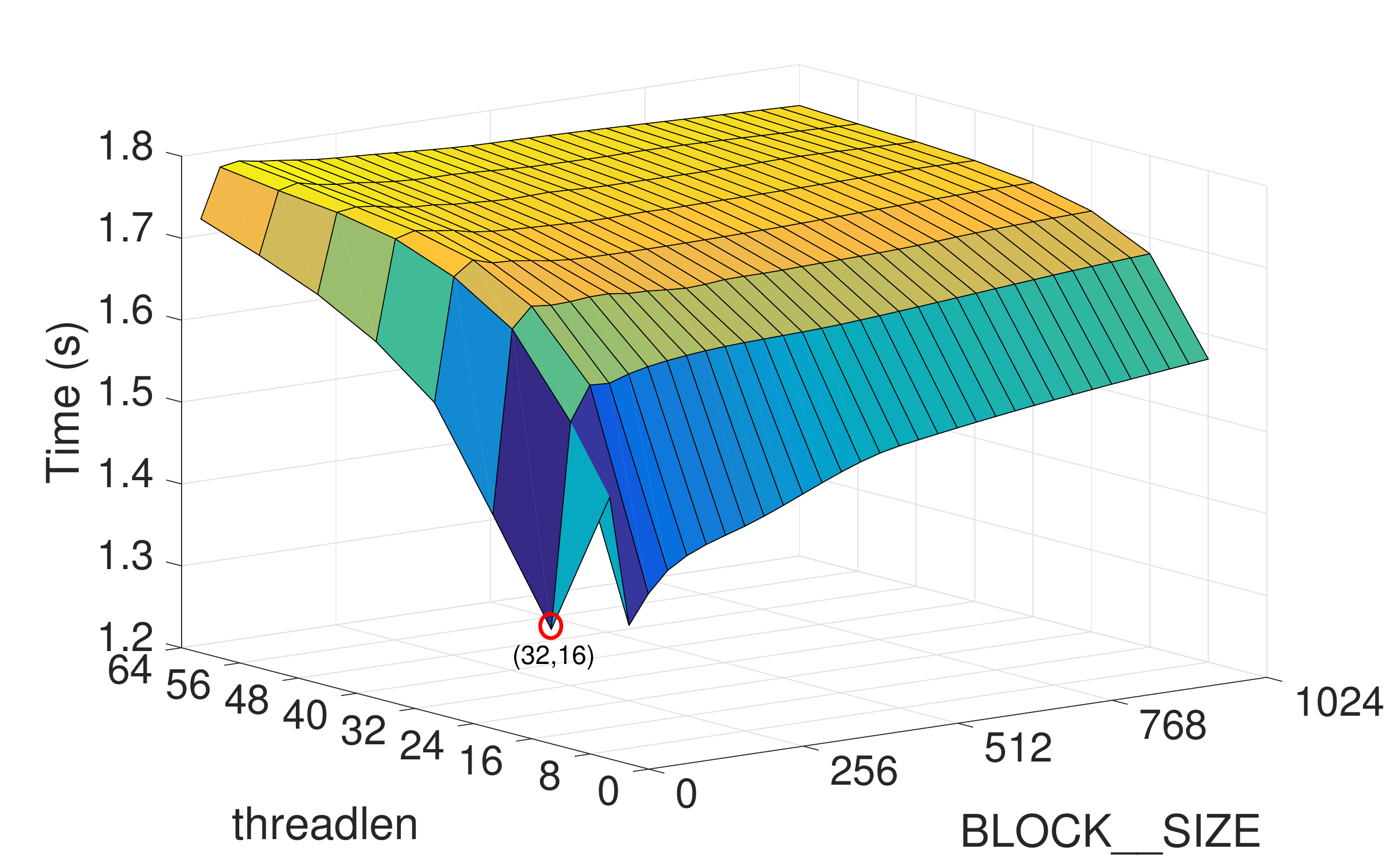} 
} 
\caption{Tuning the best parameters \textit{threadlen} and \textit{BLOCK\_SIZE} for SpMTTKRP on mode-1.} 
\label{tuning} 
\end{figure*}

Because both the sparsity pattern of the tensor and its partitioning scheme will impact the memory footprint of sparse tensor operations on GPUs, we tune the \textit{threadlen} and \textit{BLOCK\_SIZE} parameters to find their best configuration.
The parameter \textit{threadlen} indicates the number of non-zeros processed by each thread and \textit{BLOCK\_SIZE} shows the number of threads inside a thread block. 
As shown in Figure \ref{tuning}, the best parameter configurations for nell1 and brainq for SpMTTKRP on mode-1 are (\textit{BLOCK\_SIZE}=128 and \textit{threadlen}=64) and (\textit{BLOCK\_SIZE}=32, \textit{threadlen}=16) respectively. The best parameter configuration for each dataset and sparse tensor operation can be found in Table \ref{tb:parameters}. These parameters are used to obtain the results for unified in the upcoming sections.

\begin{table}[!htp]
\centering
\begin{tabular}{|l|l|l|l|l|}
\hline
Operations & nell-1  & delicious  &  nell-2   &  brainq   \\
\hline
SpTTM     & (32,8)  & (512,8)    & (256,64)  & (1024,32)    \\
\hline
SpMTTKRP & (32,16) & (32,8)     & (1024,64) & (128,64) \\
\hline
\end{tabular}
\caption{The best parameters for SpTTM on mode-3 and SpMTTKRP on mode-1.}\label{tb:parameters} 
\end{table}

\subsection{Performance Results and Analysis}

\begin{figure*} 
\centering 
\subfloat[SpTTM on mode-3] { \label{ttm_speedup} 
\includegraphics[width=0.45\textwidth, height=0.3\textwidth]{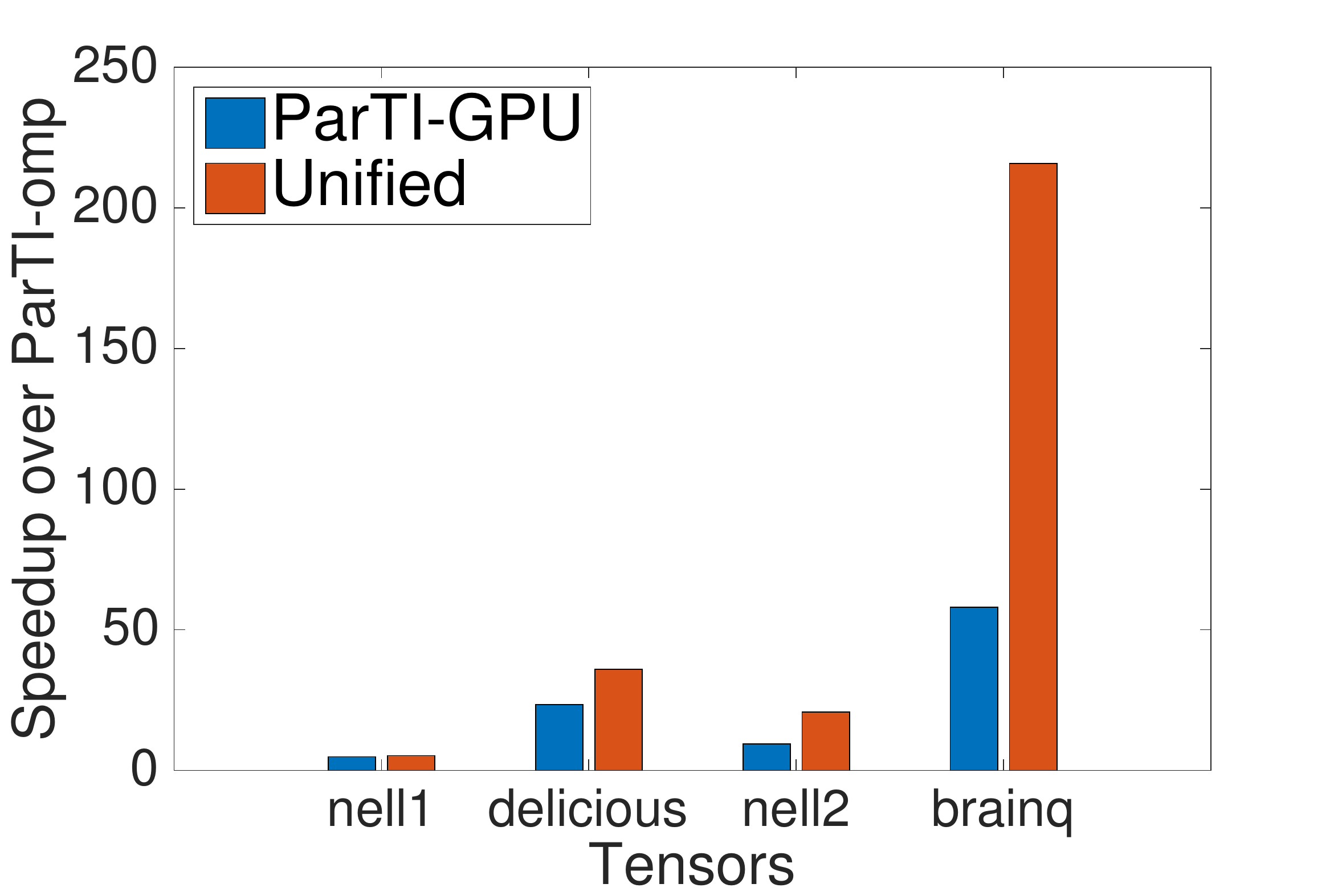} 
} 
\centering
\subfloat[SpMTTKRP on mode-1] { \label{mttkrp_speedup} 
\includegraphics[width=0.45\textwidth, height=0.3\textwidth]{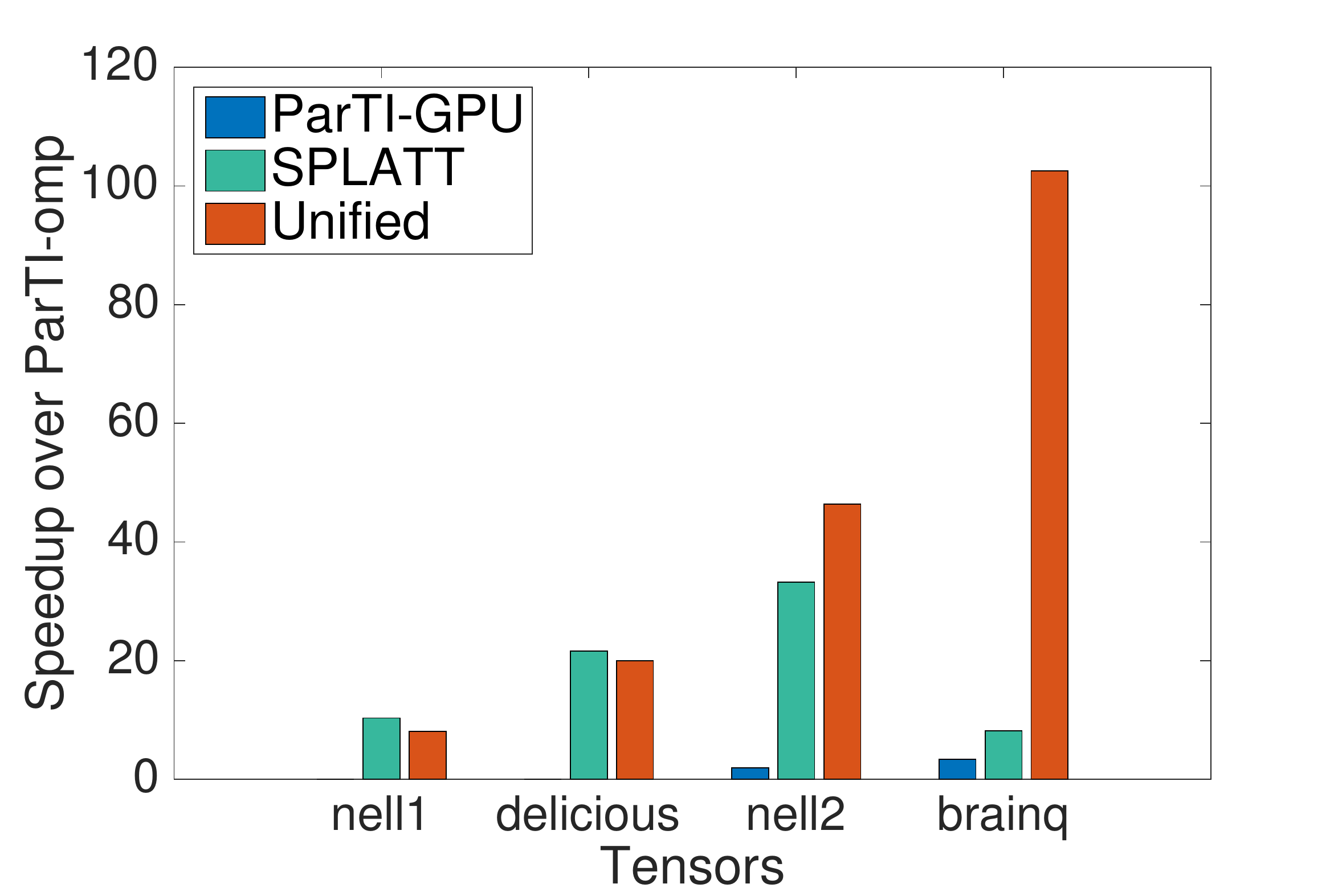} 
} 
\caption{Unified's speedup over ParTI and SPLATT (rank=16); higher is better.} 
\label{performance}
\end{figure*}

This section compares the performance of the unified method with ParTI and SPLATT. ParTI-omp and ParTI-GPU are parallel implementations of ParTI on multi-core and GPU architectures respectively. The number of columns of dense matrices in tensor computations is set to 16, which is also the rank of the tensor decomposition. Since SPLATT only supports SpMTTKRP, the performance of unified method for the SpTTM operation is only compared to ParTI. As shown in Figure \ref{ttm_speedup} compared to ParTI-omp, unified achieves $5.3\times$ (nell1) to $215.7\times $ (brainq) speedup. Compared to ParTI-GPU, Unified achieves $1.1\times$ (nell1) to $3.7\times$ (brainq) speedup. 
For the SpMTTKRP operation ParTI-GPU runs out of memory for larger tensors such as nell1 and delicious. As shown in Figure \ref{mttkrp_speedup},
compared to ParTI-omp, our proposed method achieves from $8.1\times$ (nell1) to $102.5\times$ (brainq) speedup. Compared to ParTI-GPU,  we achieve $23.7\times$ speedup on nell2 and $30.6\times$ speedup on brainq. Unified achieves a speedup of  $1.4\times$  for nell2 and $12.5\times$  for brainq compared to SPLATT.

%\textcolor{red}{More investigation and discussion of the results are necessary. In particular, the biggest gains of they approach happen for "brainq", which is a relatively small and dense (30\% sparsity) tensor. Could this be the effect of initial overheads in the other solutions? Note that SPLATT does better than Unified for the larger tensors.}

%\textcolor{red}{Figure 8 only discusses the "brainq" dataset, which is the smallest and densest of the 4 tensors. Do things change for the larger datasets?}

The experiments show the unified method achieves best performance for the brainq dataset and does not perform well for the nell1 dataset. %\textcolor{black}{Since the speedup achieved by unified method decreases when density of dataset decreases : tensor becomes sparser, the density is one core factor determining performance gain achieved by unified method. For example, unified method works better for delicious (140M) than nell1 (144M) which is close to nell1's size, the reason is that these two dataset differs in the density.} 
nell1 is extremely sparse with a density of $9.1e-13(nell1)$ while brainq is the densest tensor in our dataset. The performance of tensor operations on GPUs tend not to be good for very sparse tensors because the non-zero elements processed by one warp will need access to columns of the dense factor matrices that are located far apart based on the indices of the product mode. To resolve this issue, we use a read-only data cache to cache accesses to the dense matrices. However, if the indices in the product mode vary to a large extent, cache hit rates will decrease. Thus while unified does not perform well for nell1 it performs better for brainq (density: $2.9e-01$), nell-2 (density: $2.5e-05$), and delicious (density: $6.1e-12$).

% \begin{figure*} 
% \centering 
% \subfloat[SpTTM on mode-3] { \label{ttm_speedup} 
% \includegraphics[width=0.24\textwidth]{ttm_speedup.eps} 
% } 
% \centering
% \subfloat[SpMTTKRP on mode-1] { \label{mttkrp_speedup} 
% \includegraphics[width=0.24\textwidth]{mttkrp_speedup.eps} 
% } 
% \subfloat[SpTTM] { \label{mode-ttm} 
% \includegraphics[width=0.24\textwidth]{ttm_mode.eps} 
% } 
% \centering
% \subfloat[SpMTTKRP] { \label{mode-mttkrp} 
% \includegraphics[width=0.24\textwidth]{mttkrp_mode.eps} 
% } 
% \caption{(\ref{ttm_speedup}), (\ref{mttkrp_speedup}):  Unified Method's Speedup over ParTI and SPLATT (rank=16).\label{performance}. (\ref{mode-ttm}), (\ref{mode-mttkrp}): Mode Behavior of different implementations (rank=16) on the "brainq" dataset.} 
% \label{mode} 0.36 0.24
% \end{figure*}

\subsection{Mode Behavior}
  
\begin{figure*} 
\centering 
\subfloat[SpTTM] { \label{mode-ttm} 
\includegraphics[width=0.45\textwidth, height=0.3\textwidth]{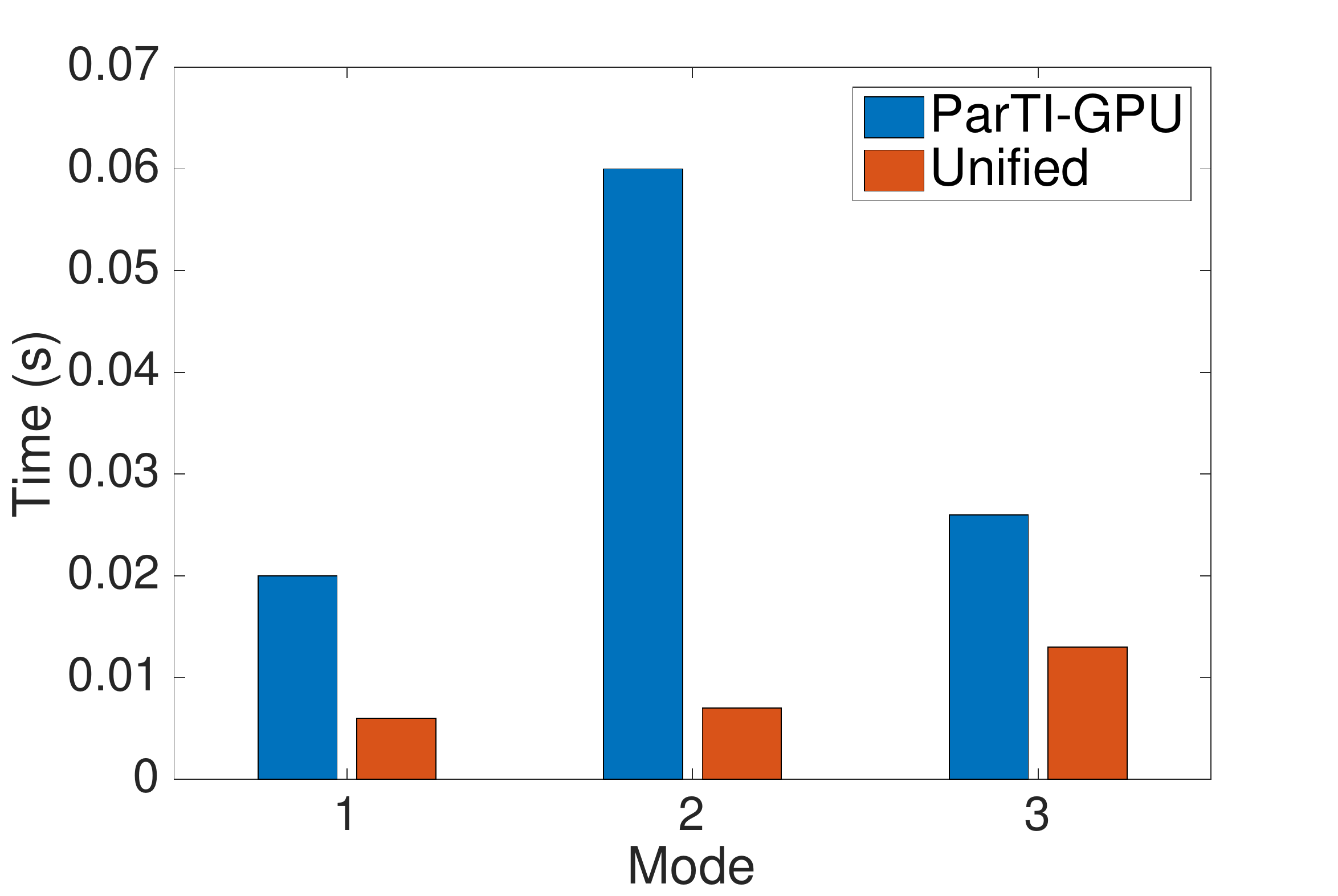} 
} 
\centering
\subfloat[SpMTTKRP] { \label{mode-mttkrp} 
\includegraphics[width=0.45\textwidth, height=0.3\textwidth]{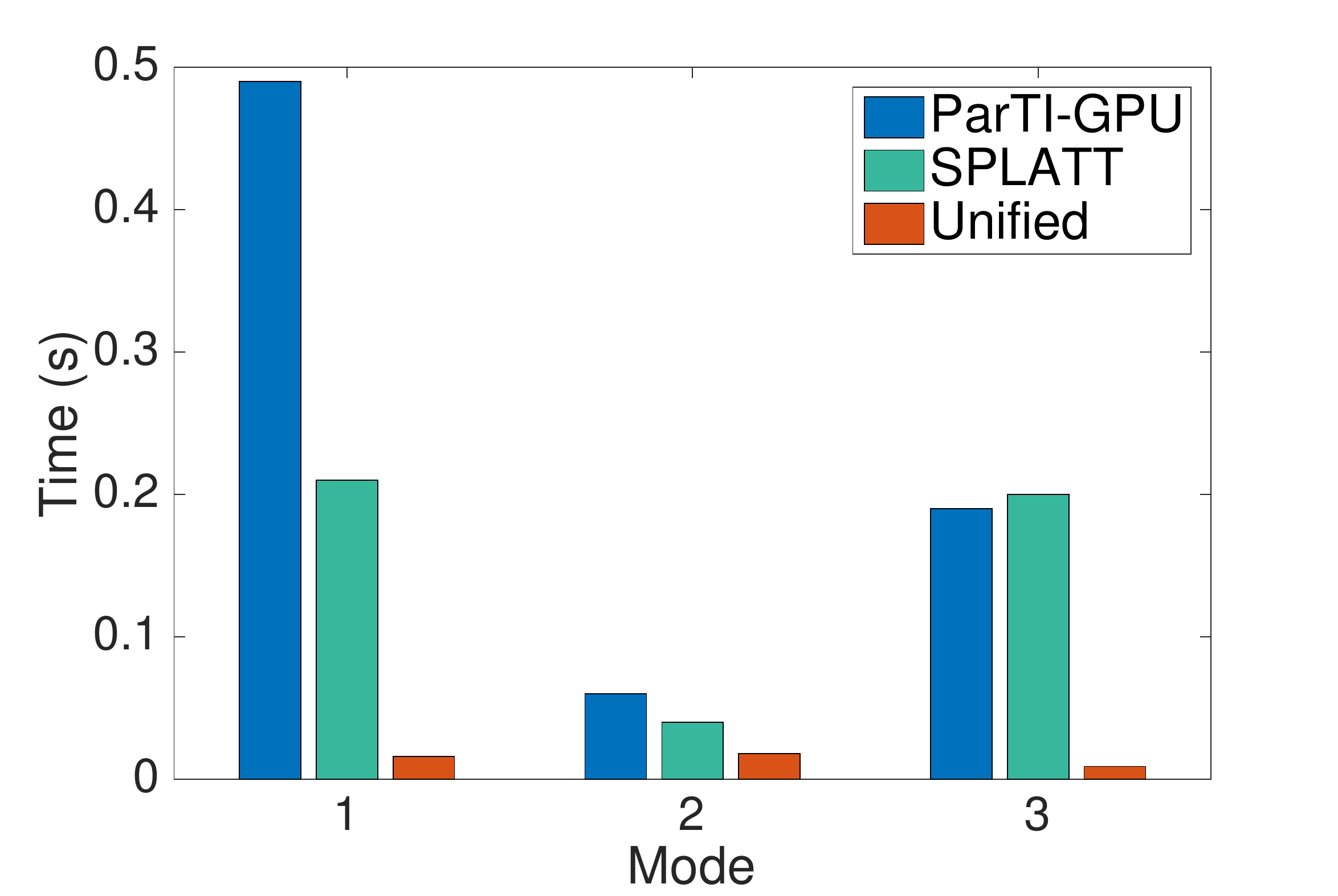} 
} 
\caption{Mode behavior of different implementations (rank=16) on the "brainq" datasets; lower is better.} 
\label{mode}
\end{figure*}

The experiments in this section demonstrate that the performance of the unified method does not depend on the mode being operated on and performs relativity the same mainly because it is based on the F-COO format.
Since brainq is one of the ``oddly" shaped tensors in our dataset with $60 \times 70K \times9$ dimensions it is used to examine the performance of unified on different modes. 
As shown in Figure \ref{mode}, for both SpTTM or SpMTTKRP, unlike ParTI-GPU and SPLATT, the running time of
unified method  remains relatively the same for different modes. 

Unified performs well for all modes because it uses the F-COO format to partition and parallelize based on the tensor non-zeros. However, because ParTI-GPU operates on fibers of the sparse tensor, its performance changes for each mode. For example, when computing SpTTM on mode-2 of brainq, ParTI only launches up to $540$ threads to perform computations on fibers in parallel and thus does not efficiently use the GPU resources.  SPLATT organizes the sparse tensor as a tree. Thus, parallelizing computations for different modes requires operating on different levels of the tree which changes the level of parallelization and the memory access patterns.

\subsection{Rank Behavior}
\begin{figure}
  \centering
    \includegraphics[width=0.45\textwidth,height=0.32\textwidth]{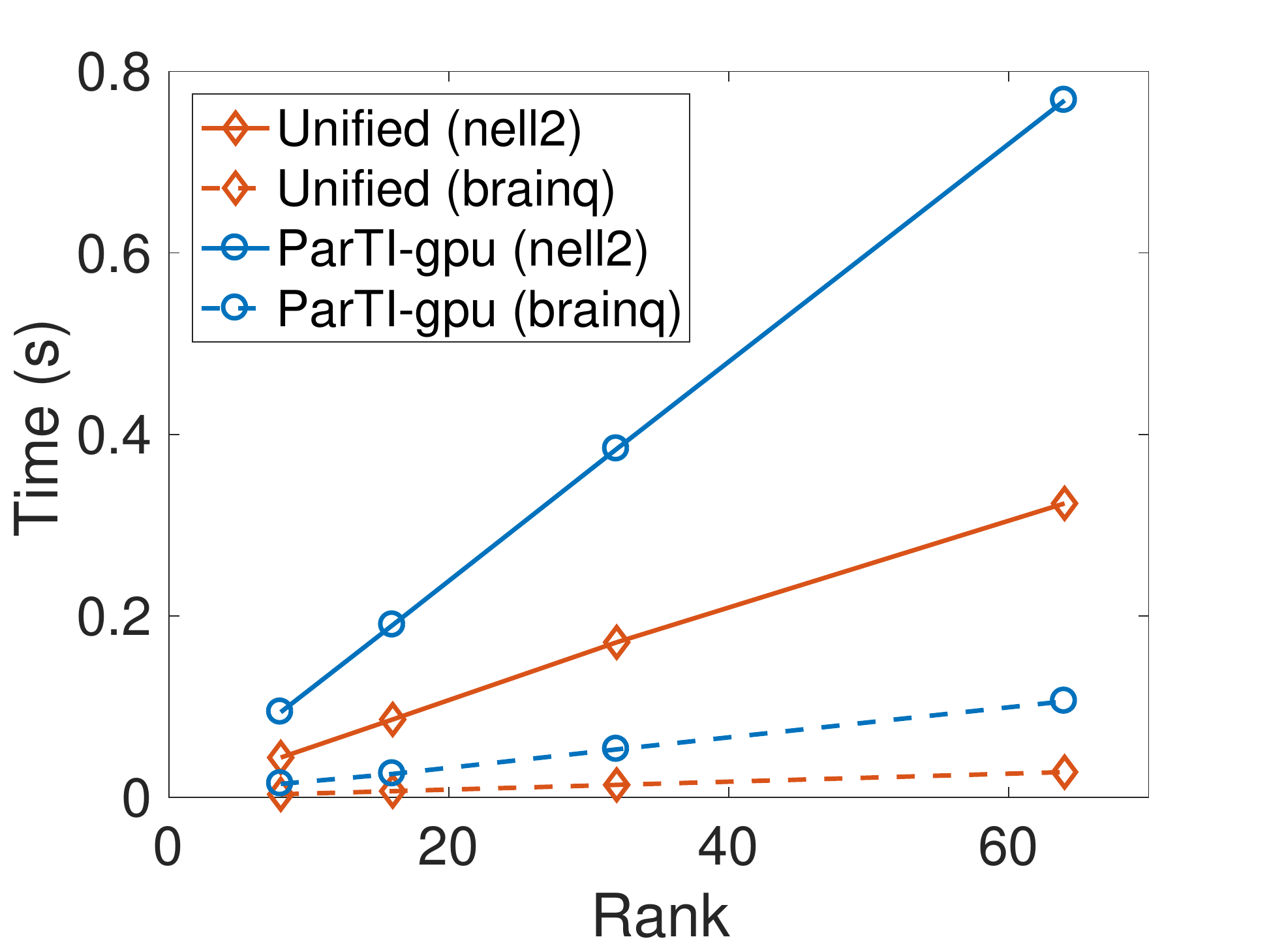}
  \caption{The execution time of unified and ParTI for SpTTM with different rank sizes for the "brainq" and "nell2" datasets; lower is better.}
  \label{rank}
\end{figure}

% We also test our method on different ranks  to check whether the performance of our method is affected by the varies of rank. 
As discussed in Section IV-D, the performance of unified scales well when the number of columns in the dense factor matrices is increased, i.e. increasing the rank of the tensor decomposition. We ran unified for different ranks ($8,16,32,64$). As demonstrated in \cite{li2016optimizing}, when the rank increases the resulting tensor or matrix from the sparse tensor operations becomes larger, thus, we only test rank behavior for the two smallest tensors “brainq” and “nell2”.
As shown in the Figure \ref{rank}, when the rank varies from 8 to 64, the execution time of ParTI increases at a faster rate compared to unified. The speedup of unified over ParTI-GPU for brainq varies from $3.7\times$ to $4.3\times$ and the speedup of unified over ParTI-GPU for nell1 varies from $2.1\times$ to $2.4\times$.  %Thus, the performance of our method will not be affected negatively by value of rank.   

\subsection{Memory Consumption on GPUs}
\begin{figure}
  \centering
  \includegraphics[width=0.45\textwidth,height=0.3\textwidth]{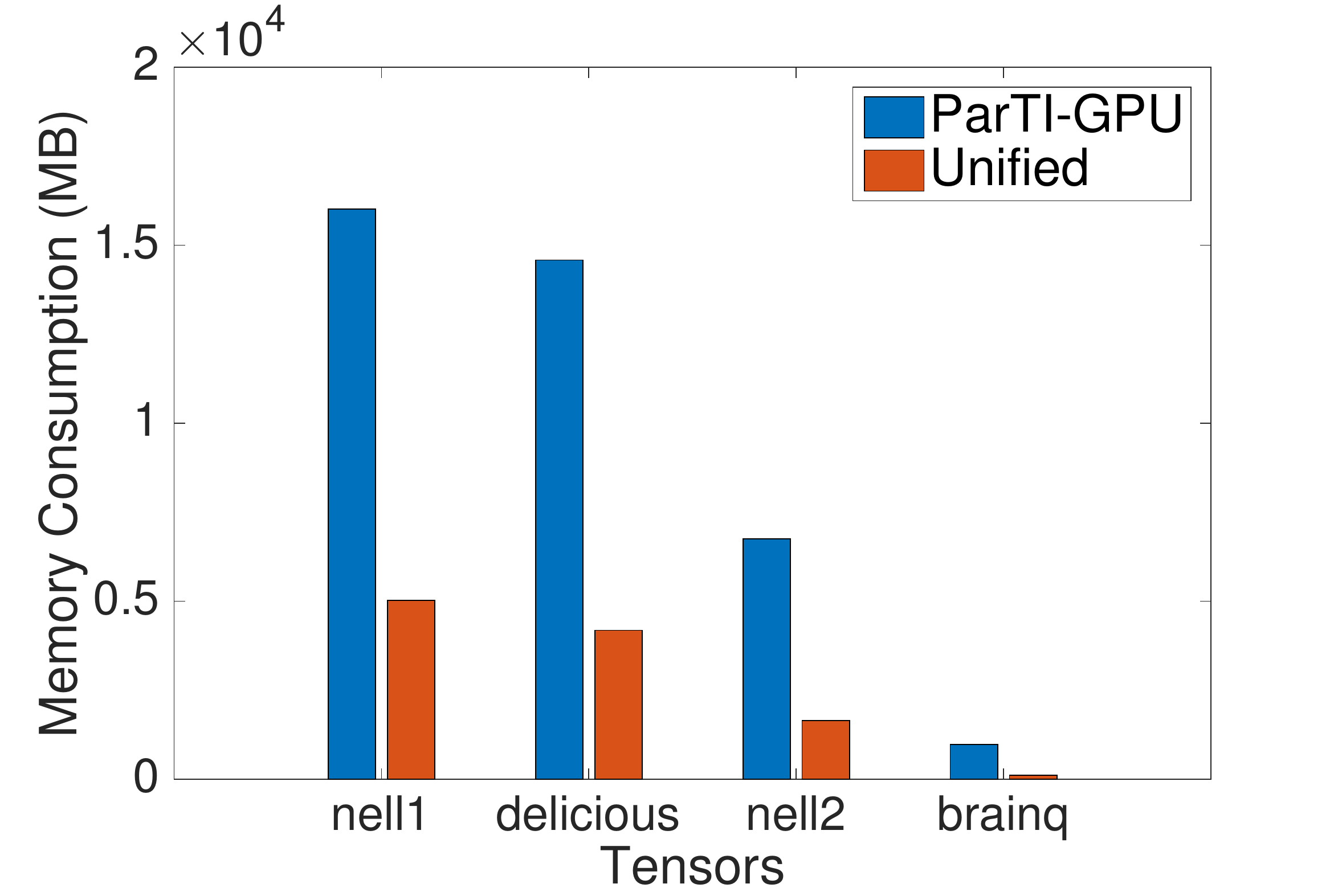}
  \caption{The GPU global memory consumption of unified vs. ParTI for SpMTTKRP on mode-1; lower is better. }
  \label{memory}
\end{figure}

SpTTM does not generate intermediate data, thus, the memory consumption of unified and ParTI is nearly the same. However, ParTI does generate intermediate data for SpMTTKRP. 
Since ParTI runs out of memory when operating on the larger tensor datasets  we computed the memory consumption by hand for nell1 and delicious based on ParTI's open source code. The memory consumption for the other two datasets is measured by executing the code. As shown in Figure \ref{memory},
compared to ParTI-GPU, our method reduces the memory consumption by $68.6\%$  for nell1 and $88.6\%$  for brainq because of the one-shot computations. 

\subsection{ CANDECOMP/PARAFAC (CP) Decomposition on GPUs}

% \begin{table*}\label{tb:cp}
% \begin{tabular}{|c|c|cccc|}
% \hline
% Dataset & \multicolumn{1}{|c|}{SPLATT}  & \multicolumn{4}{|c|}{Unified Method} \\
% % \cline{2-3} \cline{4-8}
%         & Time  & Time   & mode-1 MTTKRP & mode-2 MTTKRP &mode-2 MTTKRP  \\ 
% \hline
% brainq &  0.45  & 0.0302 $(14.9\times) $  & 0.0085 & 0.0062 & 0.0086  \\  
% nell2 &  0.70 &  0.2388 $(2.9\times)$  & 0.0547 & 0.0728 & 0.0652  \\ 
% \hline
% \hline
% \end{tabular} 
% \centering
% \caption{The performance result of unified method and SPLATT. The result includes time per iteration of SPLATT and our unified method, the decomposition rank is 8. This table also include the detailed time of SpMTTKRP on different modes per iteration}
% \end{table*}

\begin{figure}
  \centering
    \includegraphics[width=0.45\textwidth,height=0.3\textwidth]{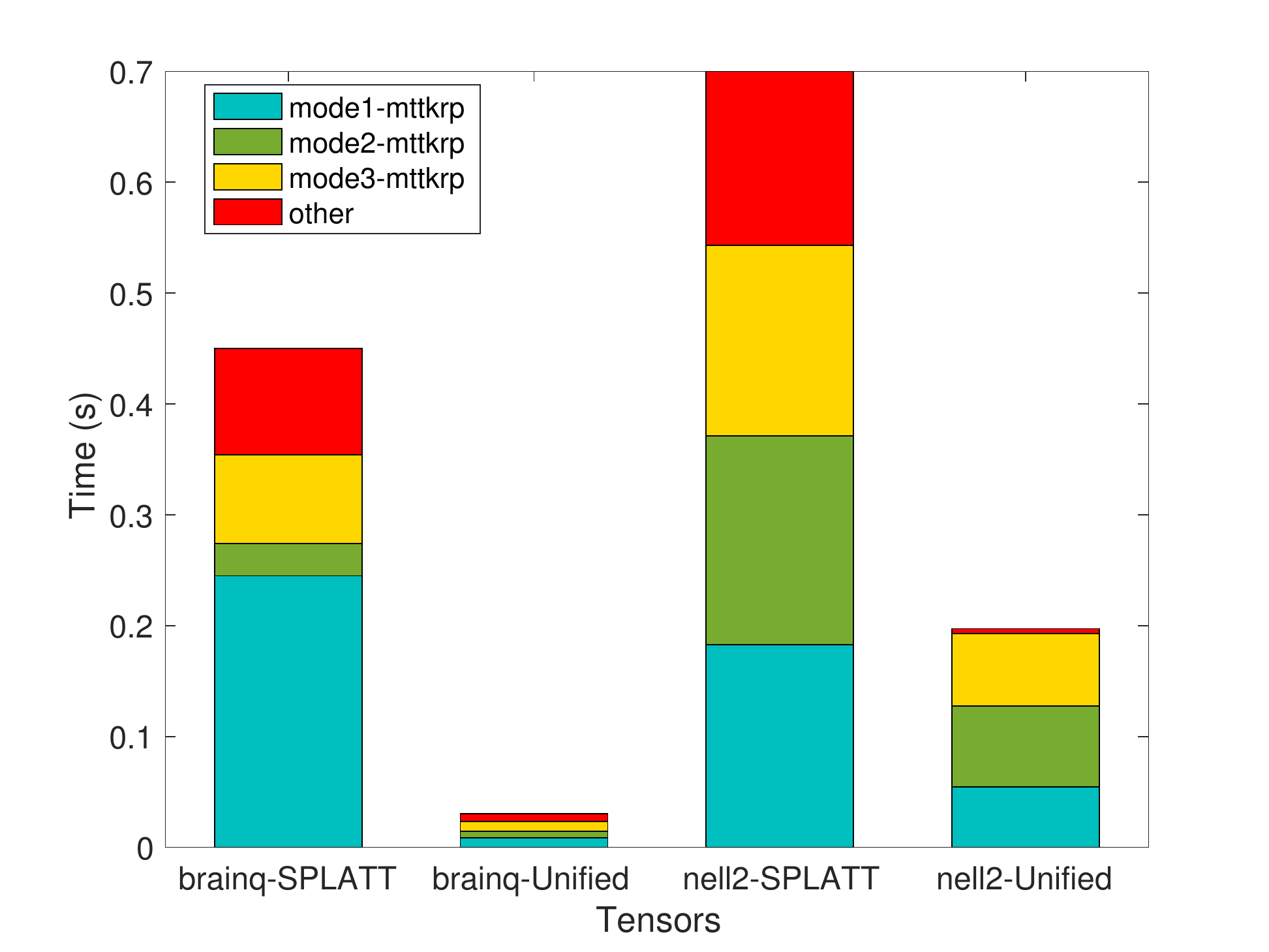}
  \caption{ Running time of unified method and SPLATT performing CP decomposition on "brainq" and "nell2" datasets; lower is better.}
  \label{cp}
\end{figure}

To the best of our knowledge, this work provides the first implementation for the CP decomposition on GPUs. Our implementation creates two GPU streams where one stream performs SpMTTKRP on different modes and the other performs matrix operations including matrix multiplication and matrix inversion using the CUBLAS library \cite{nvidia2008cublas}. As a result, the computations performed by separate streams would be overlapped automatically when possible. Because a single-GPU memory can not store all the tensor data for the CP decomposition we provide results for brainq and nell2. For larger datasets, multiple GPU cards can be used. 
As shown in Figure \ref{cp}, most of execution time for CP decomposition is spent on the SpMTTKRP operation and unlike SPLATT, in unified computations are well-balanced between operations for both tensors.

%\textcolor{purple}{ what are you trying to show the reader by giving CP results? You need to first bring this paragraph into context, read my comments below.} 

{% We present the benchmark results for CP decomposition that we discuss in the background section. 
We compare the performance of SPLATT with our unified method for CP decomposition. ParTI doesn't support CP on GPUs. The rank of the tensor decomposition is fixed to 8 to represent the low-rank property of the tensor decomposition. Another important reason is that one of the dimensions in brainq $(60\times70k\times9)$ is 9 and ranks larger than 9 will create a deficient matrix in the tensor decomposition algorithm. From Figure \ref{cp}, the unified method achieves $14.9\times$ and $2.9\times$ speedup over SPLATT for brainq and nell2 respectively. As shown in Figure \ref{cp}, most of time in our implementation is spent on performing SpMTTKRP operations.%, the time per iteration in CP decomposition is roughly three times a single SpMTTKRP.% Hence, there is almost no overhead to apply our unified method for sparse tensor operations in a actual scenario of tensor algorithm
}

%\textcolor{blue}{Bangtian: This is current result, I will see whether we can achieve more by profiling and analysis the code}
%\textcolor{red}{The objective of putting CP results is to show we do not need mode or data transfer. In other words for your CP results here is what I think we should report: 1) a table that shows the CP convergence in Unified CP is the same as a solid CP implementation (maybe Matlab or SPATT). In that table or graph we show that the total iterations in Unified CP do not change for specific convergence criteria compared to the stable Matlab or SPLATT implementation. 2) We show that the time per iteration in Unified-CP is roughly three times a single MTTKRP this way the reader believes our claim that no time is spent on mode conversion or data transfers. This might not even be needed if we show exact convergence. 3) The performance has to be compared to ParTI's GPU, I recall you mentioned they do have a repo and you can make their code run for GPUs. If so, compare to them. If not, we have to very clearly say this is the first GPU implementation of CP, and our objective is to show that unified easily extends to decomposition methods if the tensors fit on the GPU. 4) The reviewer will like to see why brainq speedups are less then nell. Also what happened to the other tensors.  }

\section{Conclusion}
This paper proposes a unified sparse storage format and parallel algorithms for sparse tensor operations on GPUs. The tensor modes for different operations are encoded into unified to deliver highly-efficient implementations of sparse tensor operations. Unified is used across different tensor operations and also accelerates the execution of complete tensor-based algorithms.   Unified's performance is not sensitive to mode changes in tensor methods, scales well with rank updates, and reduces memory footprints and storage costs on GPUs. Several techniques are used to further improve the performance of unified on GPUs such as the segmented scan method, kernel fusion, warp shuffle, and data reuse.  The experiments show that our unified solution significantly outperforms state-of-art implementations of sparse tensor operations on multicore CPUs and manycore GPUs. 

% Based on this view, we propose a COO-based storage format for sparse tensors performing sparse tensor operations. 
% We propose a novel unified GPU-based optimization method for parallel sparse tensor operations.  

% We introduce segmented scan primitive to sparse tensor operations.

%In the future, we aim to improve the performance of our method for extremely sparse tensors like nell1. We will extend our work to multi-GPUs to support larger tensors.
%Furthermore, we will release our work as a sparse tensor algebra library, more sparse tensor operations and algorithms will be supported in our library. 

\bibliographystyle{IEEEtran}

\bibliography{IEEEabrv,acmart.bib}

\end{document}